\documentclass[iop]{emulateapj}
\usepackage{mathptmx}
\usepackage{natbib}
\usepackage{units}
\usepackage{graphicx}
\usepackage{url}
 \usepackage{color}
\bibpunct{(}{)}{;}{a}{}{,}
\usepackage[colorlinks=true, urlcolor=blue, citecolor=blue, linkcolor=blue]{hyperref}

\newcommand{\be}{\begin{equation}}
\newcommand{\ee}{\end{equation}}

\newcommand{\bifrost}{{\textsl{Bifrost}}}

\newcommand{\ci}{\ion{C}{1}}
\newcommand{\cii}{\ion{C}{2}}
\newcommand{\ciii}{\ion{C}{3}}
\newcommand{\civ}{\ion{C}{4}}
\newcommand{\cv}{\ion{C}{5}}
\newcommand{\mgii}{\ion{Mg}{2}}

\newcommand{\caii}{\ion{Ca}{2}}
\newcommand{\ciib}{\ion{C}{2}~\ensuremath{133.5\,{\rm nm}}}
\newcommand{\ciia}{\ion{C}{2}~\ensuremath{133.4\,{\rm nm}}}

\newcommand{\is}{\ensuremath{\!=\!}}
\def\edta#1{{#1}}
\def\edt#1{{#1}}

\begin{document}

\title{%
 The formation of  {\it IRIS} diagnostics
 \\V. A quintessential model atom of \mbox{C II} and general formation
  properties of the \mbox{C II} lines at $133.5\,{\rm nm}$  
}%
\author{Bhavna Rathore}
\author{Mats Carlsson}
\affil{
Institute of
Theoretical Astrophysics, University of Oslo, P.O. Box 1029
Blindern, N-0315 Oslo, Norway
}
\email{bhavna.rathore@astro.uio.no}
\email{mats.carlsson@astro.uio.no}

\begin{abstract}
%Some of the strongest lines in the solar UV spectrum come from singly ionized carbon, at wavelengths of 
%$133.4 - 133.6\,{\rm nm}$. 
The \ciib\ lines are important observables for the 
NASA/SMEX mission {\it Interface Region Imaging Spectrograph }({\it IRIS}). 
To make 3D non-LTE radiative transfer 
computationally feasible it is crucial to have a model atom with as few levels as possible while retaining the 
main physical processes.  We here develop such a model atom and we study the general formation 
properties of the \cii\ lines. We find that a nine-level model atom of \ci-\ciii\ with the transitions treated assuming
complete frequency redistribution (CRD) suffices to describe the \ciib\ lines. 
3D scattering effects are important for the intensity in the core of the line. 
The lines are formed in the optically thick regime.
The core intensity
is formed in layers where the temperature is about 10~kK at the base of the transition region. 
The lines are 1.2--4 times wider than the atomic absorption profile due to the formation in the optically thick
regime. The smaller opacity broadening happens for single peak intensity profiles where the
chromospheric temperature is low with a steep
source function increase into the transition region, the larger 
broadening happens when there is a temperature increase from the photosphere to the low chromosphere leading to a local 
source function maximum and a double peak intensity profile with a central reversal.
Assuming optically thin formation with the standard coronal approximation leads to several errors: Neglecting 
photoionization severly underestimates the amount of \cii\ at
temperatures below 16~kK, erroneously shifts the formation from 10~kK to 25~kK and leads to too low
intensities.
\end{abstract}

\keywords{line: formation -- radiative transfer -- Sun: atmosphere -- Sun: chromosphere -- Sun: transition region}

\section{Introduction}
% Why is understanding the chromosphere important?
%   energy that heats the corona has to go through the chromosphere
%   chromosphere needs a factor of 10 more heat than the corpna
In radiative equilibrium, the temperature naturally decreases with height. It therefore came as a surprise when
it was realised that the solar outer atmosphere shows a temperature increase, from about 6000~K at the visible 
surface to millions of degrees in the corona. It is clear that the energy necessary to sustain these high 
temperatures originates in the solar convection zone but the mechanisms of transportation and dissipation of 
the energy to the outer solar atmosphere are still hotly debated.

A central role in this puzzle is played by the solar chromosphere --- the region a few thousand kilometres thick 
between the solar photosphere and the transition region and corona. This is the region where we go from a 
magnetic field being pushed around by the plasma motions in the lower parts (high value of plasma $\beta$, the
ratio of the gas pressure to the magnetic pressure) to a region where the magnetic pressure dominates. 
The energy that heats the solar corona needs to be transported through the chromosphere. The total energy 
needed to balance the radiative losses in the chromosphere is also at least ten times larger than what is needed
to sustain a hot corona
\citep{1977ARA&A..15..363W}. %Withbroe & Noyes 1977

% Chromospheric diagnostics difficult 
%   limited number of diagnostic lines available from the ground
%   UV lines needed - space
%  IRIS specifically designed to study the interface region between the photosphere and corona
In order to spectroscopically diagnose the chromosphere, we need spectral lines with high enough opacity to
have the formation there. This means resonance lines of abundant elements from the ionization state that 
dominates under chromospheric conditions or lines from excited levels of hydrogen or helium. In the optical
part of the spectrum accessible from ground based observatories we are restricted to a handful of spectral 
lines that meet these criteria: the most important being the hydrogen H-$\alpha$ line, the resonance lines and 
infrared triplet lines from singly ionized calcium and the helium 1083~nm line. 
These diagnostic lines have provided a wealth of 
information on the solar chromosphere but there are challenges in the interpretation of the observations: the H-$\alpha$ line
has a very large thermal width and is a very strongly scattering line largely decoupled from the local 
conditions
\citep{2012ApJ...749..136L}, % Leenaarts et al 2012
the calcium lines carry the chromospheric signal only in the core of a strong photospheric absorption line
and the interpretation is heavily dependent on numerical modelling
\citep{1997ApJ...481..500C} % Carlsson & Stein 1997
while the helium 1083~nm line has a complicated formation with a lower level mainly populated from
ionization from coronal radiation followed by recombination.

The ultraviolet (UV) part of the spectrum provides many more diagnostic possibilities for chromospheric
studies. There are many resonance lines from abundant elements in this part of the spectrum, 
the continuum radiation is much fainter in the UV, and the chromospheric diagnostics are intrinsically stronger. 
The recently launched NASA/SMEX mission {\it Interface Region Imaging Spectrograph (IRIS)} was especially
designed to target the solar chromosphere and transition region
\citep{2014SoPh..289.2733D}.
\textit{IRIS} provides high spectral, spatial and temporal resolution in three wavelength bands: 
$133.2 -135.8\,{\rm nm}$  and $138.9- 140.7\,{\rm nm}$    with 1.3 pm  wavelength pixels (corresponding
to $2.9 - 2.7$~km~s$^{-1}$) and $278.3 - 283.4\,{\rm nm}$ with 2.5 pm wavelength pixels (2.7~km~s$^{-1}$),
0.167 arcsec spatial pixels and 0.33 arcsec  slit width. 
%   optically thick formation
% CII lines are strong and promising
%    among IRIS main lines
%    need to investigate their formation characteristics to explore the diagnostic potential

The resonance lines from singly ionized carbon (\cii) around $133.5\,$nm are among the strongest  lines in the solar ultraviolet spectrum. There are three components in the multiplet: at $133.4532\,{\rm nm}$ and
at $133.5708\,{\rm nm}$ with a weaker blend at $133.5663\,{\rm nm}$. The large number of photons emitted in the lines makes it possible to observe at high spatial and temporal resolution and the \cii\ lines are therefore among the more important diagnostic lines for 
\textit{IRIS}. To explore this diagnostic potential it is important to find out how the lines are formed and thereby establish how the conditions in the solar outer atmosphere are encoded in the line profiles. Singly ionized carbon is the dominant ionization state throughout most of the upper chromosphere but \cii\ also exists in significant amounts up to middle transition region temperatures,
as we will show here. 

% what has been done before

The very first \cii\ solar spectrum was taken by the normal-incidence grating spectrograph from the U.S. Navel Research Laboratory (NRL) April 19, 1960 \citep{1961AnG....17..263D}. Later, the lines have been observed several times from the series of Orbiting Solar Observatory (OSO) satellites, especially OSO-4, although this instrument only obtained barely resolved \cii\ spectra \citep{1964SSRv....3..816P}.  These authors suggested that the primary emission comes from regions with a temperature of  20 kK. Later, \citet{1971SAOSR.338.....C} used OSO-4 and OSO-6
observations of the \cii\ lines to constrain the temperature
in the upper chromosphere and found the observations to be compatible with a temperature plateau 
at about 15 kK. 
\citet{1978ApJ...222..333L} % Lites, Shine & Chipman 1978
used OSO-8 observations to
constrain the temperature structure of the upper chromosphere using centre-to-limb measurements
of the \cii\ lines. Through spectrum synthesis of the lines they found them to be extremely sensitive
to the temperature and physical extent of the plateau at 20 kK proposed by 
\citet{1973ApJ...184..605V} % Vernazza et al 1978
and found the observations more compatible with a plateau at 16.5 kK with 25\% more material than
in the models by \citet{1973ApJ...184..605V}.
The OSO-8 observations reported on by \citet{1978ApJ...222..333L}  were the highest spectral resolution
observations thus far and the average quiet Sun profiles show a clear central reversal and a red emission
peak slightly brighter than the blue peak.

A detailed discussion of the formation of the \ciib\ multiplet appears in \citet{2008ApJS..175..229A}. This
study is based on a one-dimensional, time-independent semi-empirical model constructed to reproduce the
average quiet-Sun observations presented in the SUMER atlas of the extreme ultraviolet spectrum 
\citep{2001A&A...375..591C}. % Curdt et al
\citet{2013ApJ...779..155A} presented further calculations of the  \ciib\ multiplet in four semiempirical 
models representing
the faint and mean internetwork, a network lane, and bright network.

% Giants and supergiants
The \cii\ lines have also been used to diagnose other stars. 
\citet{1984ApJ...287L..43B} %{BrownCarpenter84}
used them to constrain the chromospheric temperature structure in late type giant and supergiant stars. 
%When the \cii\ lines are seen in late type spectra it indicates the existence of chromospheric activity in the stars with shallow convective envelopes. 
% F, G, K dwarfs
%The line is also prominent in F, G and K dwarfs. A strong ultraviolet \cii\ $133.5\,{\rm nm}$ line is the characteristic of free-strong objects. This is because the effect of photospheric background is very low, while the ionization potential is high enough for the line not to be affected by photospheric radiation. 
The \cii\ multiplet is  also considered as a good diagnostic to estimate stellar basal fluxes \citep{1995A&ARv...6..181S}.  A comprehensive study based on  the analysis of time series data with the Solar Ultraviolet Measurement of Emitted Radiation (SUMER) instrument on-board SOHO was presented by
\citet{2003ApJ...597.1158J}. % Judge, Carlsson & Stein
By comparing the measured intensities with synthetic data from simulations of acoustic waves they conclude that even the lowest observed levels need to have a magnetic heating component in addition to heating
by the dissipation of acoustic waves, thus questioning the common thought that the basal flux from stellar
chromospheres is accounted for by non-magnetic heating.

% what we will do here
% 
With the advent of detailed, 3D radiation magnetohydrodynamic (RMHD) models spanning from the
convection zone to the corona it has become possible to study the formation of spectral diagnostics in
an inhomogeneous and dynamic setting. Such studies provide a more extensive understanding of how
the atmospheric properties are encoded in the detailed line-profiles than the earlier studies based on 
one-dimensional semi-empirical models. In the first papers in this series of papers on the formation of
\textit{IRIS} diagnostics we have employed this approach to study the \mgii~h \& k lines
\citep{2013ApJ...772...89L, 2013ApJ...772...90L,2013ApJ...778..143P} and the \mgii\ triplet lines
\citep{2015arXiv150401733P}.
We will now present results
for the \cii\ lines. To make 3D non-LTE radiative transfer computationally feasible, it is crucial to have
a model atom with as few levels as possible while retaining the main physical processes. In this first
paper we develop such a model atom and we study the general formation properties of \edt{the intensity profiles of} the \cii\ lines.
In the next paper we will present statistical correlations between the atmospheric parameters and 
observables. In the third paper on the \cii\ lines we will use \textit{IRIS} observations to further test 
the diagnostic potential of the \cii\ lines and use the comparison of observations with the synthetic 
observables to draw conclusions on what might be missing physical ingredients in the simulations.

The layout of this paper is as follows: in Section~\ref{sec:rt} we give details on how we solve the
coupled equations of statistical equilibrium and radiative transfer, in  Section~\ref{sec:ma} we
describe the various model atmospheres we use, in Section~\ref{sec:qam} we describe the model
atom, the atomic data and the procedure that ends up with a quintessential model atom.  In Section~\ref{sec:res} we present the general formation properties of the \cii\ lines, in Section~\ref{sec:sunspot}
we discuss the formation in a Sunspot atmosphere and we end with conclusions and discussion
in Section~\ref{sec:cd}.

\section{Radiative transfer}
\label{sec:rt}

For the detailed analysis of  the \cii\ lines we solve the statistical equilibrium equations in non-LTE using three codes. For the study of our atomic model, basic radiative transfer processes and the simplifications of the model atom we use the 1D code MULTI  \citep{1986UppOR..33.....C}. In order to study 3D effects we use the full 3D radiative transfer code MULTI3D \citep{2009ASPC..415...87L}. To study the importance of partial redistribution (PRD) effects and effects of calculating background continua with several elements in non-LTE simultaneously we use the code RH \citep{2001ApJ...557..389U}.
MULTI  solves the non-LTE radiative transfer problem in semi-infinite, plane parallel one dimensional atmospheres with prescribed macroscopic velocity fields. Complete frequency redistribution (CRD) is assumed for all bound-bound transitions and the background scattering is considered to be coherent. Line transitions that overlap in frequency are not treated self consistently (which is of relevance here since one of the \cii\ lines is a blend with two components). Version 2.3 of MULTI includes the local approximate operator of \citep{1986JQSRT..35..431O}, Ng acceleration \citep{1974JChPh..61.2680N}, collisional-radiative switching  \citep{1988A&A...192..279H} and the opacities are taken from the Uppsala Opacity Package \citep{Gustafsson1973}. The background continuum opacity is normally calculated assuming LTE for the relevant atoms and ions but it is possible to iteratively include non-LTE populations for the background opacity. 

The full three dimensional radiative transfer code MULTI3D is not only superior in that it treats the non-LTE problem in full 3D but the treatment of overlapping lines and continua is self-consistent and not based on iteration. However, the computations are naturally more computationally intensive in 3D which is why we use the 1D version for experimentation with large model atoms.
We assume CRD also for the computations with MULTI3D.
%and for the computationally intensive computation of response functions, see section~\ref{sec:resp}.

RH is another non-LTE code capable of treating PRD in the lines and calculating overlapping continua by solving for several elements in non-LTE simultaneously. This code is therefore used for the tests of the significance of PRD and for the detailed continuum calculations.

\section{Model atmospheres}
\label{sec:ma}

We use a number of different model atmospheres in this work. We start by studying the atomic processes important for the lines we are interested in by using the often used semi-empirical  VAL3C model atmosphere \citep{1981ApJS...45..635V}. Once we have arrived at a simplified atomic model that preserves the important characteristics of the more extended atomic model, we employ a 3D model atmosphere to test the simplified model atom over a wider parameter range and to see the sensitivity of the line formation to atmospheric properties. To study radiative transfer effects in 3D, we employ the full 3D solution. 
%We also study 1D solutions with input atmospheres taken from columns along a cut in the 3D model atmosphere. 
The 3D atmosphere is taken from a 3D simulation with the 
 \bifrost\ code \citep{2011A&A...531A.154G}.  \textit{Bifrost} solves the equations of radiation magnetohydrodynamics on a staggered Cartesian grid. Detailed radiative transfer is included through the multi-group opacity approach 
 \citep{1982A&A...107....1N} % Nordlund 1982
modified to take into account scattering 
\citep{2000ApJ...536..465S}. % Skartlien 2000
Radiative losses in the chromosphere 
are calculated in non-LTE using simplified recipes 
\citep{2012A&A...539A..39C} %Carlsson & Leenaarts 2011
that are based on detailed 1D full non-LTE radiative transfer
simulations.  Non-equilibrium ionization of hydrogen is included following the
description of
\citet{2007A&A...473..625L}. % Leenaarts et al (2007).
The simulation we use encompasses $24 \times 24 \times 16.8 $ Mm on the Sun discretised onto a   $504 \times 504 \times 496$ grid. 
Vertically, the computational volume extends from 2.4 Mm below to 14.4 Mm above average optical depth unity at 500 nm (which is the zero-point of our height-scale), and thus covers the solar atmosphere from the upper convection zone, photosphere, chromosphere to the lower corona. 
Horizontally, the grid is equidistant with a  spacing of 48 km. 
The z-axis is not equidistant. 
It has grid spacing of 19 km between $z\!=\!-1$ and $z\!=\!5$ Mm, whereas the spacing increases towards the lower and upper boundaries to a maximum of 98 km at the coronal boundary.  
Magnetic fields are introduced into a relaxed hydrodynamic simulation by specifying the vertical magnetic field  at the bottom boundary with a potential field extrapolation producing the magnetic field throughout the 3D box. 
In our particular simulation we specify two concentrations of opposite polarities separated by 8 Mm. The potential field is quickly swept to the intergranular lanes and slowly builds up a set of loops between the opposite polarities. We have chosen a snapshot at a time 3850s after the magnetic field was introduced, well after initial transients have passed through the box and the heating of the atmosphere through the work of the convection on the magnetic field has come to a quasi-steady state. 
This simulation is described in detail in 
\citet{en024048} % Carlsson et al simulation description
and is the same as used in 
\citet{2012ApJ...749..136L}, %Leenaarts et al, H-alpha formation
\citet{2012ApJ...758L..43S}, %Stepan et al, the Hanl\'e effect of Ly-$\alpha$
\citet{2013ApJ...764L..11D} %de la Cruz Rodriguez et al, funny profiles
and for the first papers in this series of papers on the formation of IRIS diagnostics
\citep{2013ApJ...772...89L, 2013ApJ...772...90L,2013ApJ...778..143P,2015arXiv150401733P}.

\begin{figure}[hbtp]
 \includegraphics[width=\columnwidth]{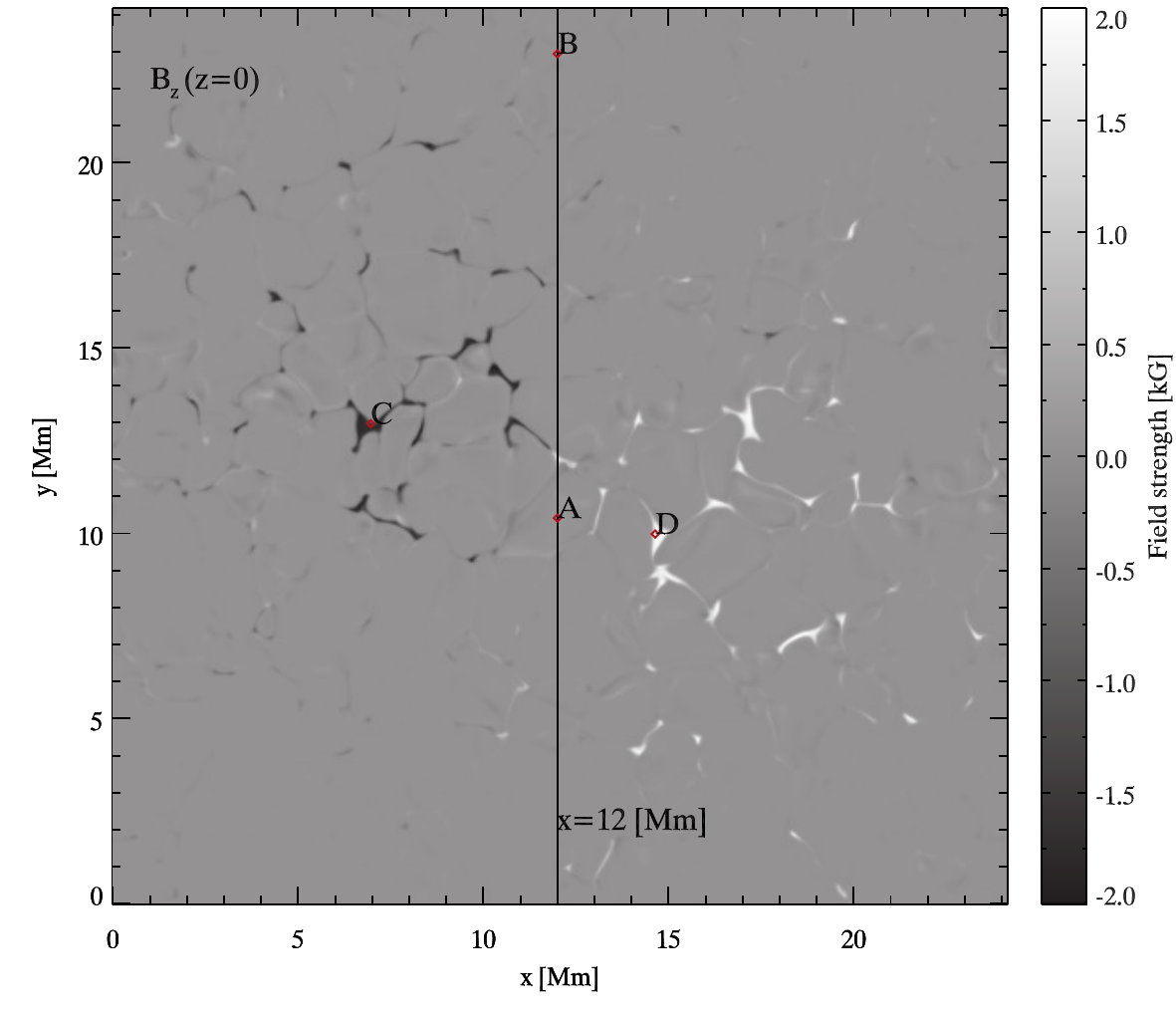}
 \caption[]{\label{int_mag} 
Vertical magnetic field of the \bifrost\  atmosphere at $z\!=\!0$ height. The vertical line shows the position of the 2D cut used and the letters A-D mark columns chosen for detailed study in section~\ref{sec:cf}.  }
\end{figure}

The vertical magnetic field strength in the photosphere is shown in Figure~\ref{int_mag}. 
The two 
opposite magnetic field polarity patches are clearly seen. We will illustrate formation properties along a slice through the 
model at $x\is12\,$Mm, marked with a vertical line, and at four different positions, marked with letters A-D in Figure~\ref{int_mag}.
%The emergent intensity for the \cii\ (133.5 nm ) multiplet clearly shows the loops connecting the two polarities (right  panel of  Figure~\ref{int_mag}). 
%A full 3D non-LTE calculation for our simplified \cii\ atomic model takes 2 hour wall clock time running simultaneously on 512 cores on a Cray XE6 computer. 

To study the formation of the \cii~lines also in a more active atmosphere, we employ a semi-empirical atmosphere of a sunspot umbra. The standard model SPOTM of 
\citet{1986ApJ...306..284M} % Maltby et al 1986
has an emphasis on the temperature structure of the photosphere up to the temperature minimum region and does not extend high enough to cover the formation region of the \cii~lines. We therefore use the semi-empirical model of 
\citet{1982ApJS...49..293L} % Lites & Skumanich 1982
that is based on OSO-8 observations of the chromospheric lines of hydrogen (Ly-$\alpha$, Ly-$\beta$), \mgii~(k,h), \caii~(K,H) and the transition region \civ~$1548\,{\rm nm}$ resonance line.

%%%%%%%%%%%%%%%%%%%%%%%%%%%%%%%%%%%%%  %===================================41========================================
%===========================================================================
%%%%%%%%%%%%%%%%%%%%%%%%%%%%%%%%%%%%%

\begin{figure}[hbtp]
 \includegraphics[width=\columnwidth]{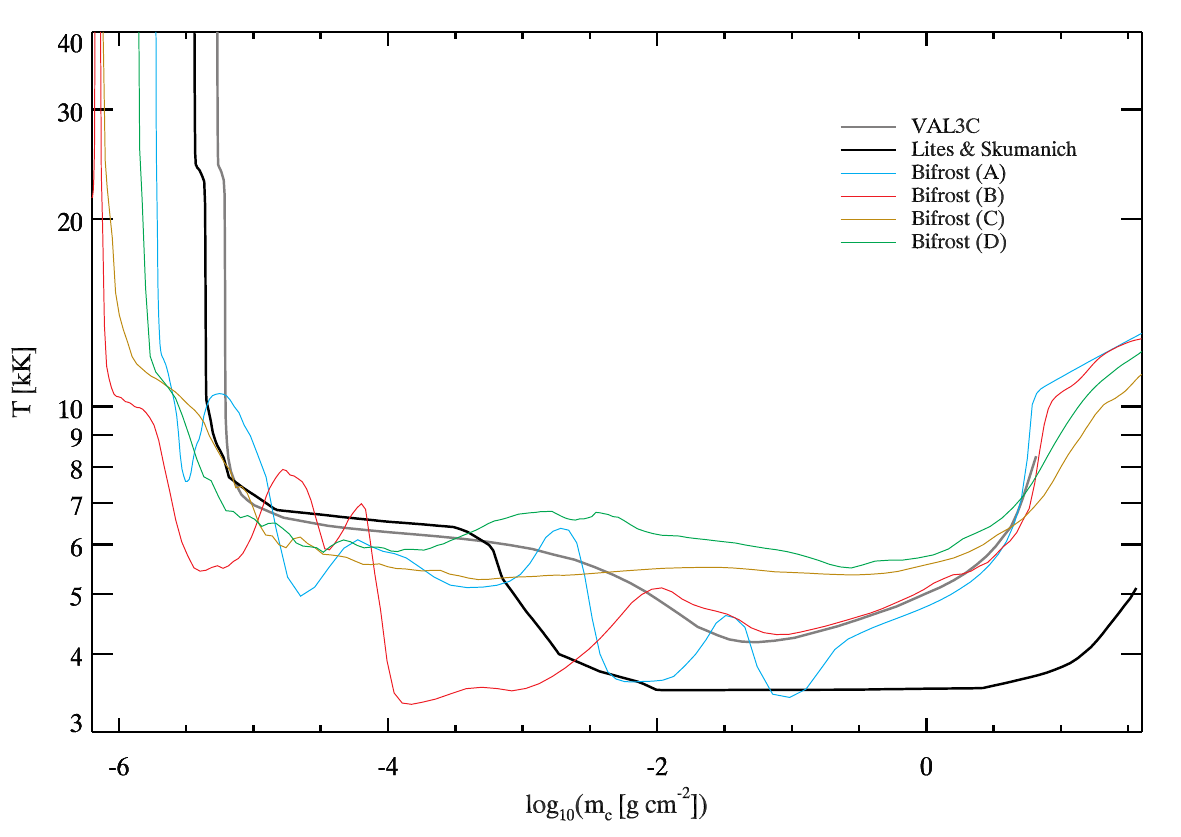}
 \caption[]{\label{fig:atmos_mod} 
 Temperature as function of logarithmic column mass for atmospheric models in this paper: 
 The quiet Sun model VAL3C  (thick grey), 
 the umbral model of \citet{1982ApJS...49..293L} (thick black),
 and four columns from the \bifrost\ snapshot as marked in Figure~\ref{int_mag}: A (blue), B (red), C (brown) and D (green).
}
\end{figure}

The temperature structures of the two semi-empirical models and the four columns in the \bifrost\ snapshot are shown in 
Figure~\ref{fig:atmos_mod}. Both the semi-empirical models have an artificial temperature plateau at 24~kK in order to
get enough flux in the hydrogen Lyman-$\alpha$ line. The umbral model has this plateau and the transition region at
a slightly smaller column mass than the quiet Sun model VAL3C. The transition region in the \bifrost\ model is typically at an
even smaller column mass.

%MB too much white space around figure
%%%%%%%%%%%%%%%%%%%%%%%%%%%%%%%%%%%%%  %===================================41========================================
%===========================================================================
%%%%%%%%%%%%%%%%%%%%%%%%%%%%%%%%%%%%%

\section{Quintessential model atom}
\label{sec:qam}

\begin{figure}[hbtp]
  \includegraphics[width=\columnwidth]{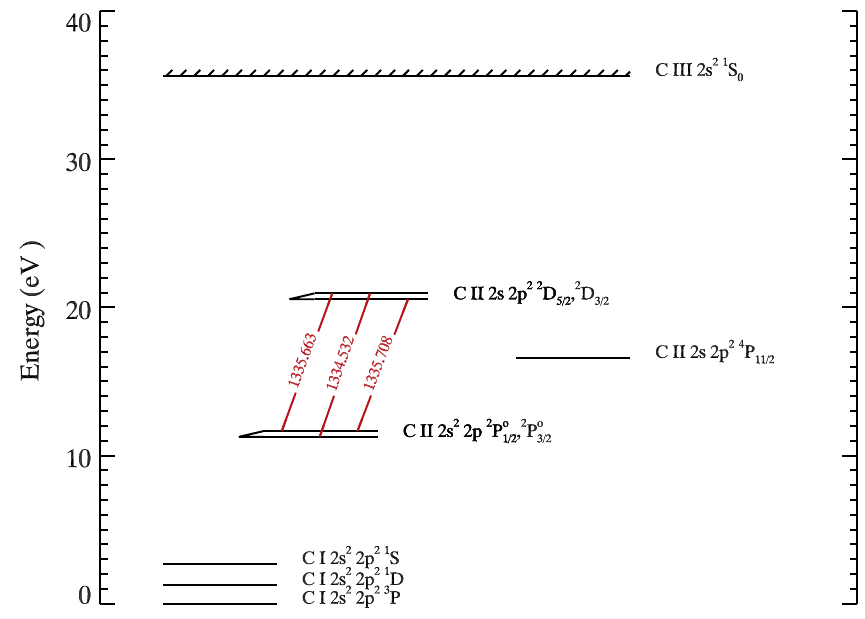}
  \caption[]{\label{tdc9} 
  \cii\ atomic model with 9 levels. The \cii\ multiplet of special interest here consists of the three allowed transitions between the $2s\,2p^2\,^2\!D$ and the $2s^2\,2p\,^2\!P^o$ terms.
  }
\end{figure}

\begin{figure}[hbtp]
  \includegraphics[width=\columnwidth]{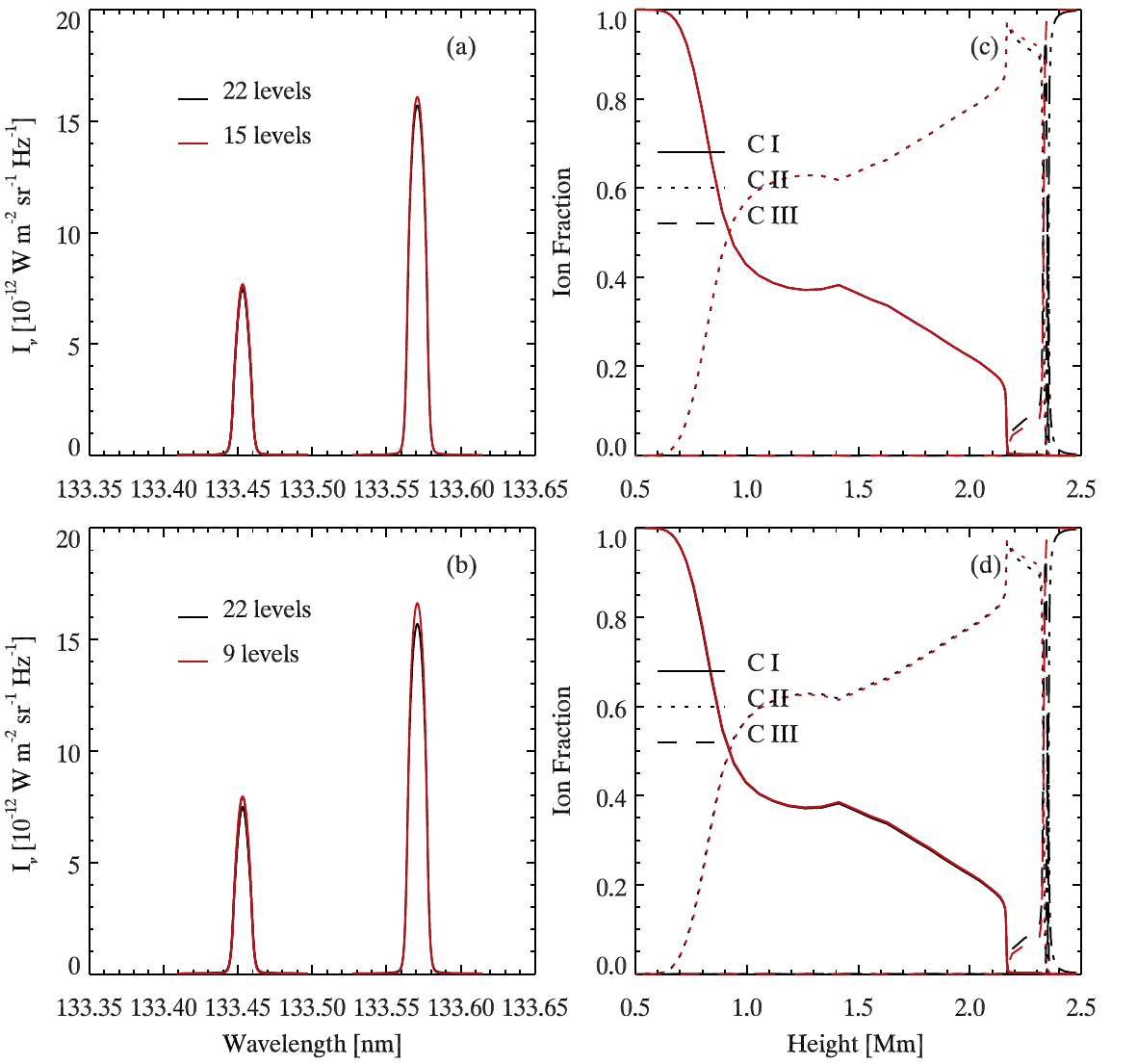} 
  \caption[]{\label{intinf} 
Intensities and ionization fractions resulting from the original 22-level atom and the simplified 15-level atom 
(panels (a) and (c)) and the quintessential 9-level atom (panels (b) and (d)). Results for the
reference model are shown in black and for the 15/9-level models in red.
The VAL3C model atmosphere was used and the weak blend at 133.566  nm was neglected. }
\end{figure}

One of the objectives of the current work is to arrive at a simplified model atom that includes the non-LTE processes important for the formation of the \edt{intensity profiles of the} \ciib\
lines but is small enough to enable large scale 3D non-LTE simulations. To that end we start with a more complex model atom, the one used 
by \citet{2003ApJ...597.1158J}.  This atomic model contains a total of 22 levels: 8 levels of \ci, 6 levels of \cii, 4 levels of \ciii, 3 levels of \civ\ and the ground 
state of \cv. We simplify this model by removing and merging levels following the procedure described in \citet{2008ApJ...682.1376B}. % Bard & Carlsson 2008
At each step we solve the non-LTE problem and compare the intensities in the \ciib\ lines and the ionization fraction as function of depth with that of the full model atom.
As a first step we remove all levels above the ground state of \ciii. This results in a model atom of 15 levels. In the next step we remove levels not important for the ionization balance and the levels not directly coupled to the lower and upper levels of the \ciib\ lines. The quintessential model atom we arrive at has a total of 9 levels, see
Figure~\ref{tdc9}. The lines we are interested in are the unblended component at $133.4532\,{\rm nm}$ (hereafter called the $133.4\,{\rm nm}$ line) and the stronger
$133.5708\,{\rm nm}$ line that has a weaker blend at $133.5663\,{\rm nm}$, hereafter collectively referred to as the $133.5\,{\rm nm}$ line. 
A comparison between the results for the VAL3C atmosphere based on the 22-level model atom and the 15 and 9-level atoms is given in Figure~\ref{intinf}. 
Both the intensities and ionization structure
are almost identical calculated from the quintessential 9-level model atom as from the full 22-level atom.
This is also true for the full \bifrost\ atmosphere --- on average the quintessential 9-level model has 5\% larger
peak intensity in the \cii\ lines but less than 5\% of the columns show a difference of more than 15\% and
the largest difference is 27\% compared with the full 22-level atom.

We have also made experiments with a more complete \ci\ model atom including highly excited levels to get recombination channels right
and including the photoionizing radiation from the hydrogen Lyman-$\alpha$ line. This influences the \ci/\cii\ balance at the low temperature end
but has negligible influence on the \ciib\ lines, see Section~\ref{sec:ionbal}.

\edta{We stress that our model atom has been constructed for the specific purpose of modeling the intensity of the \ciib\ lines. If other lines are of interest or if polarization is to be calculated, this model atom might need to be extended to include additional levels.}

%%%%%%%%%%%%%%%%%%%%%%%%%%%%%%%%%%%%%  
%===========================================================================
%===========================================================================
%%%%%%%%%%%%%%%%%%%%%%%%%%%%%%%%%%%%%  

%%%%%%%%%%%%%%%%%%%%%%%%%%%%%%%%%%%%%%===========================================================================
%===========================================================================
%%%%%%%%%%%%%%%%%%%%%%%%%%%%%%%%%%%%%  

%%%%%%%%%%%%%
\subsection{Atomic data}
In the 9-level atomic model level energies for \ci\ are from the NIST\footnote[1]{http://aeldata.nist.gov/nist\_atomic\_spectra.html} online database, 
for \cii\ from \citet{moortable} %moore table for C,N,O.. book
and the \ciii\ level is from \citet{1978aelr.book.....M}. The transition probabilities are from the NIST database. Detailed photoionization data (from \ci\ to \cii) are from the TOPBASE\footnote[2]{http://cdsarc.u-strasbg.fr} database of the OPACITY project. The bound-bound collisional rates for \cii\ are from 
\citet{2008A&A...486..629T}. %Tayal et al 2008 
The bound-bound collisional rates for neutral carbon are determined using the impact parameter method of  \citet{1962PPS....79.1105S}. %seaton 1962
As we will see later, the intensity ratio between the multiplet components depends on the relative population of the upper levels. As
the energy difference between the levels of the $2p\,^2\!D$ term is only 2.5 cm$^{-1}$ (0.008 eV), collisions with neutral hydrogen may be important.
For these collisions we use data from \citet{1968ApJ...152..701B} %Bahcall & Wolf 1968
with $P(i\rightarrow f)=4/10$ (Barklem 2014, private communication).
%The collisional rates between the closely coupled levels $^{2}\!P^{o}_{1/2}$ and $^{2}\!P^{o}_{3/2}$ are from  \citet{1992ApJS...80..425B}.

In our simulations the total recombination rates include the radiative rates, dielectronic recombination rates and three body recombination rates. The radiative recombination rates are either calculated from the detailed photoionization cross-sections (recombination into \ci) or as tabulated by 
\citet{1982ApJS...48...95S}. %Shull & van Steenberg 1982
The three body recombination rates are from \citet{1985A&AS...60..425A} %{Arnaud+Rothenflug1985}.
and the dielectronic recombination rates are from 
\citet{2004A&A...420..775A} (\cii\ $\rightarrow$ \ci) and
\citet{2003A&A...412..597C} (\ciii\ $\rightarrow$
\cii).
We set the carbon abundance to 8.43  \citep{2009ARA&A..47..481A} on the usual logarithmic scale where the abundance of hydrogen is 12.

\subsection{Continuum intensity} \label{sec:bkgo}

\begin{figure}[hbtp]
  \includegraphics[width=\columnwidth]{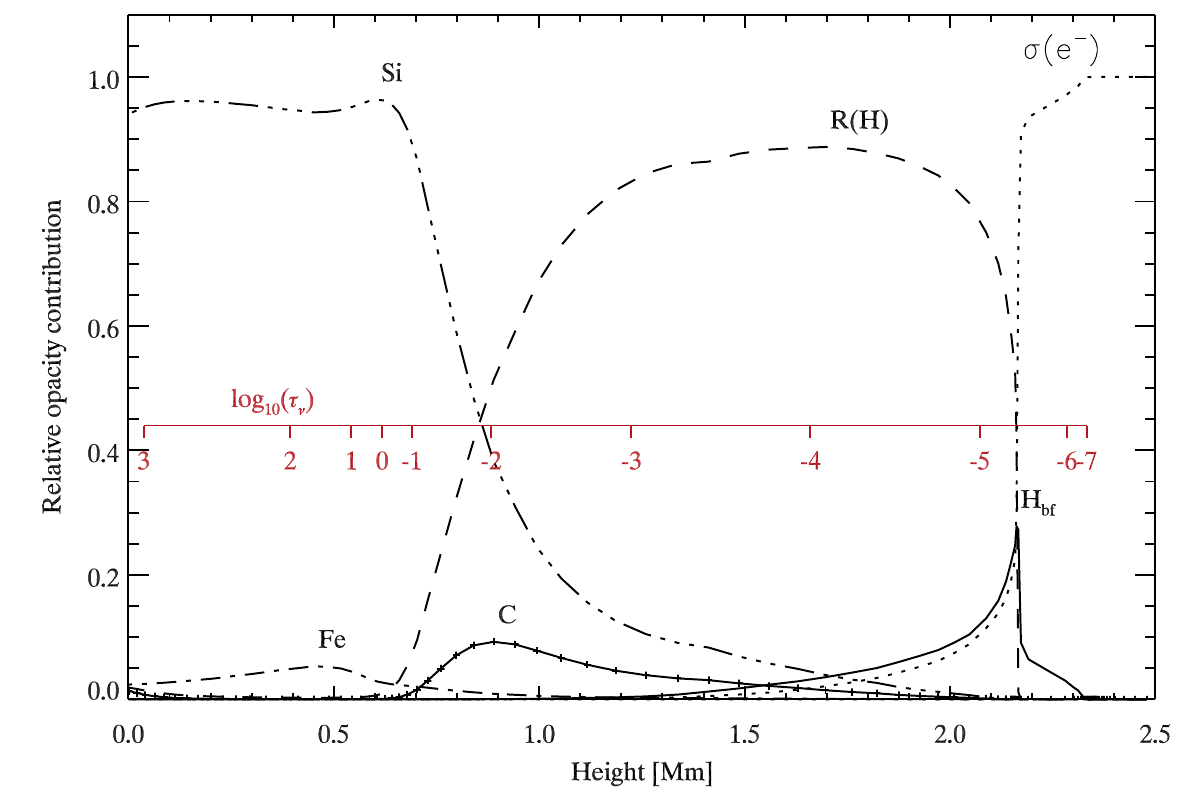}
    \caption[]{\label{bkgop} 
 Relative contribution to the background opacity at $133.5\,$nm as a function of height in the VAL3C atmosphere. 
 The source of the opacity is bound-free opacity from Si, C, Fe and H and electron scattering ($\sigma(e^-)$) and
 Rayleigh scattering on neutral hydrogen (R(H)). The secondary height scale shown in red is the 
 logarithm of the continuum optical depth at $133.5\,$nm. Hydrogen, carbon and silicon have all been treated in non-LTE
 while iron was treated in LTE.

  }
\end{figure}

\begin{figure}[hbtp] 
  \includegraphics[width=\columnwidth]{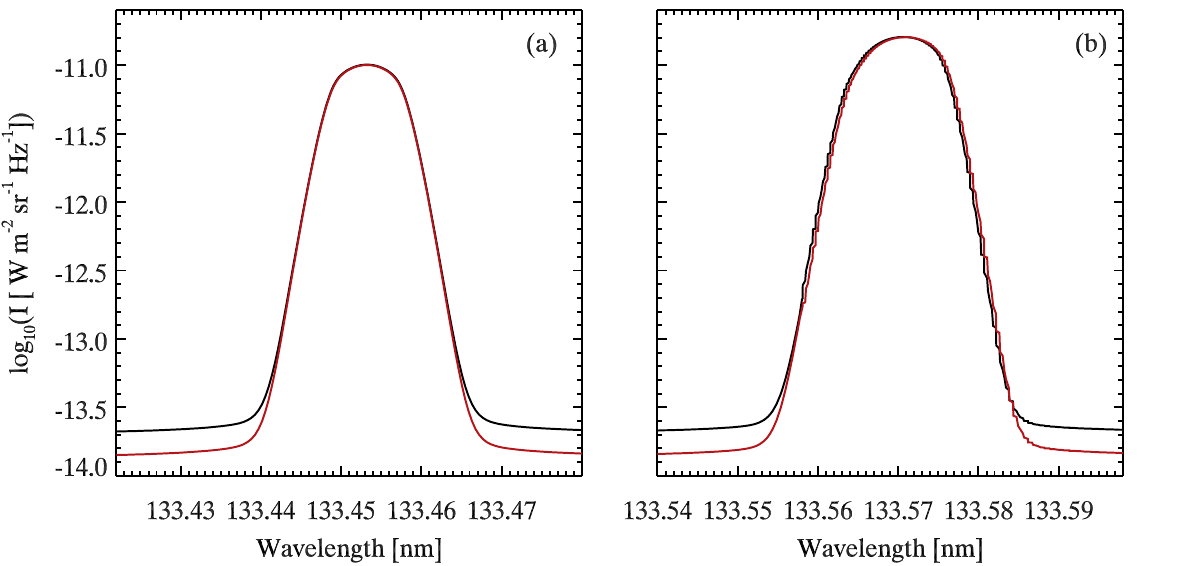}
  \caption[]{\label{si_bkg} 
  \cii\ line intensity for the \ciia\ (left) and \ciib\ (right) lines, when assuming non-LTE Si background opacity (red) and LTE Si background opacity (black). The VAL3C atmosphere was used for the comparison. Note the logarithmic intensity scale.
   }
\end{figure}

In addition to the opacity from the multiplet lines we are interested in, we have a background opacity that varies slowly with wavelength. At $133.5\,$ nm under solar chromospheric conditions, this background opacity is dominated by bound-free transitions in silicon, carbon, hydrogen and iron with electron scattering taking over as the dominant background opacity source in the upper chromosphere and corona, see Figure~\ref{bkgop}. The carbon background opacity comes from bound-free transitions from the levels $2p^2\,^{1}\!D$ and $2p^2\,^{1}\!S$ while the ground term $2p^2\,^{3}\!P$ has a bound-free edge at $110\,$nm and does not contribute. Note that the energy difference from the \ion{S}{1} levels $3p^4\,^{1}\!D$ and $3p^4\,^{1}\!S$ and the ground state of \ion{S}{2} places the bound-free edges longward of our carbon multiplet and these transitions would then contribute to the background opacity. However, these levels ionize to an excited state of \ion{S}{2} placing the bound-free edges shortward of our multiplet and sulphur therefore does not contribute significantly to the background opacity. Hydrogen, carbon and silicon are all out of LTE in the chromosphere so a proper calculation of the continuum intensity needs to treat all three elements in non-LTE simultaneously. Our carbon model atom includes the singlet levels of \ci\ so that contribution is in non-LTE automatically. Hydrogen non-LTE level populations are also provided in the atmospheric models used here so the only remaining problem is the silicon contribution. To include silicon in non-LTE simultaneously with carbon would be computationally costly. This is of course necessary if one is interested in getting the continuum intensity correct. However, in cases where the continuum intensity is very much lower than the carbon multiplet intensity, the correct treatment of the background continuum may be unimportant for getting the correct multiplet intensity. We tested this by comparing the full treatment with one where silicon is assumed to be in LTE for the calculation of the background opacity, see Figure~\ref{si_bkg}. It is clear that the continuum intensity is severely overestimated when silicon is treated in LTE but that such a simplified treatment does not significantly affect the line intensity.

%%%%%%%%%%%%%%%%%%%%%%%%%%%%%%%%%%%%%
%===========================================================================

%===========================================================================
%%%%%%%%%%%%%%%%%%%%%%%%%%%%%%%%%%%%%  

%%%%%%%%%%%%%%%%%%%%%%%%%%%%%%%%%%%%%
%===========================================================================

\subsection {Ionization balance}\label{sec:ionbal}

The formation of the \cii-lines is to a large extent set by the ionization balance -- where in the atmosphere is the \cii\ ionization stage dominant? This in turn is set by the ionization and recombination rates between \cii\ and \ciii\ and by the rates between \ci\ and \cii. 
The balance between \ci\ and \cii\ is dominated by photoionization and radiative recombination. 
To get this balance right, \edta{we must} include the levels that dominate the photoionization (normally the ground levels and levels with low excitation) and highly excited levels that dominate in the recombination channels. The photoionization of \ci\ is dominated by ionization from the $2p^2\,^3\!P$ term with edges around 110~nm and from the $2p^2\,^1\!D$ level with
photoionization edges close to 123.9 nm. For the latter, photoionization from hydrogen Lyman-$\alpha$ radiation may play an important role. This complicates the computational task enormously since the proper solution involves a large number of \ci\
energy levels and the simultaneous solution of the hydrogen non-LTE problem with the Lyman-$\alpha$ line treated in partial frequency redistribution. We have made test calculations with such a setup and find that by neglecting the photoionization from
Lyman-$\alpha$ and all levels above the ground term and the two first excited levels of \ci, \edta{the amount of \cii\ 
at temperatures below 10~kK is indeed underestimated} but the effect on the \cii\ lines is minimal \edta{(intensity changes by less
than 5\%)}.

\begin{figure}[hbtp]
  \includegraphics[width=\columnwidth]{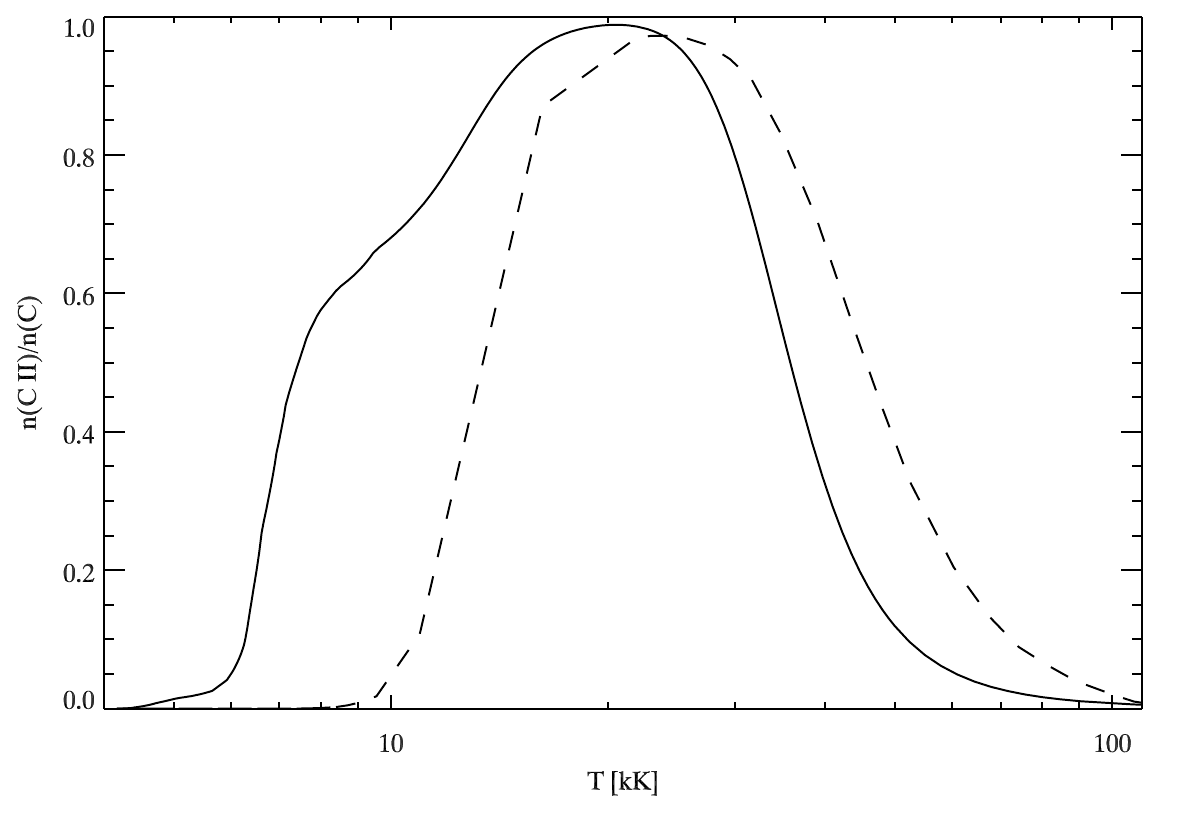}
  \caption[]{\label{mulch_ioneq} 
  \cii\ ionization fraction in the VAL3C model atmosphere using the adopted atomic parameters (solid), neglecting photoionization from \ci\ to \cii\ and also neglecting density dependent dielectronic recombination (=default Chianti ionization balance) (dashed). Note the logarithmic temperature scale.
  }
 \end{figure}

In the classical coronal ionization equilibrium approximation, photoionization is neglected and the ionization rate is assumed to be dominated by collisional ionization. This leads to a severe underestimate of the amount of \cii\ below a temperature of 20 kK under solar chromospheric conditions, as is clearly shown in Figure~\ref{mulch_ioneq}. 

At the high temperature end, the ionization fraction of \cii\ is determined by collisional ionization and dielectronic recombination. 
Part of the dielectronic recombination is to high energy levels and there may be collisional ionization taking place before the electron cascades down to lower energy levels through allowed radiative transitions. A correction in the dielectronic recombination rates is made following \citet{1972MNRAS.158..255S} %summers 1972 
as implemented by \citet{NCAR_OpenSky_The_HAO_Spectral_Diagnostic_Package_for_Emitted_Radiation_(HAOS-DIPER)_Reference_Guide_(Version_1.0)} by including a  density dependent factor in the dielectronic recombination coefficient. This decreases the recombination rate and we get a smaller \cii-fraction at the high temperature range than if the dielectronic recombinations to high energy \cii-levels are all assumed to end up with radiative cascades to the ground state, see Figure~\ref{mulch_ioneq}.
Neglecting \edta{Summer's correction factor for dielectronic recombination, the \cii\ ionization fraction does not fall 
below 10\% until a temperature of 75 kK is reached. Including this factor}, the fraction of \cii\ drops rapidly at a temperature of 25 kK and we have an ionization fraction below 10\% already at 50 kK. 

%%%%%%%%%%%%%%%%%%%%%%%%%%%%%%%%%%%%%
%===========================================================================
%===========================================================================
%%%%%%%%%%%%%%%%%%%%%%%%%%%%%%%%%%%%%

\begin{figure}[hbtp]
  \includegraphics[width=\columnwidth]{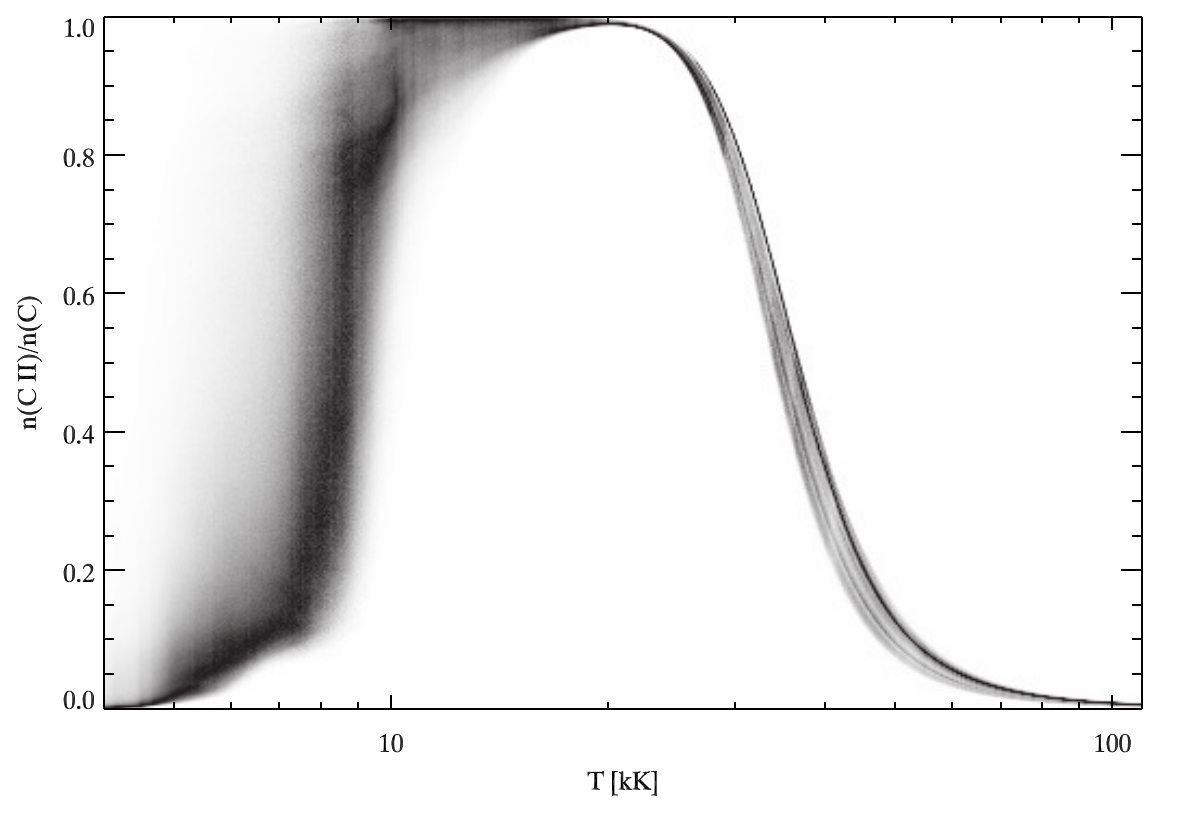}
  \caption[]{\label{inf3d} 
Probability density function of the \cii\ ionization fraction as function of temperature in the \bifrost\ simulation. In each temperature bin the PDF 
has been normalized to the largest value for increased visibility. Note the logarithmic temperature scale. }
 \end{figure}

The radiative and dielectronic recombination rates are a function of temperature and linearly dependent on electron density apart from the above mentioned density \edta{correction} of the dielectronic recombination rate. Therefore, if the ionization is dominated by collisional ionization (also a function of temperature and linear in electron density) and three body recombination is negligible, the ionization balance is a function of temperature only. In Figure~\ref{inf3d} we show the fraction of \cii\ in the full \bifrost\ snapshot (504x504x496 grid points). It is clear that in the high temperature end we are close to having an ionization balance only dependent on temperature. The spread is caused by three-body recombination and the density dependent dielectric recombination. At the low temperature end, the ionization is dominated by photoionization and the ionization fraction is not a single valued function of temperature. There is also an extra spread caused by the 3D radiation field.

%%%%%%%%%%%%%%%%%%%%%%%%%%%%%%%%%%%%%
%===========================================================================
%===========================================================================
%%%%%%%%%%%%%%%%%%%%%%%%%%%%%%%%%%%%%

\subsection {Effect of Partial frequency Redistribution}

\begin{figure}[hbtp]
  \includegraphics[width=\columnwidth]{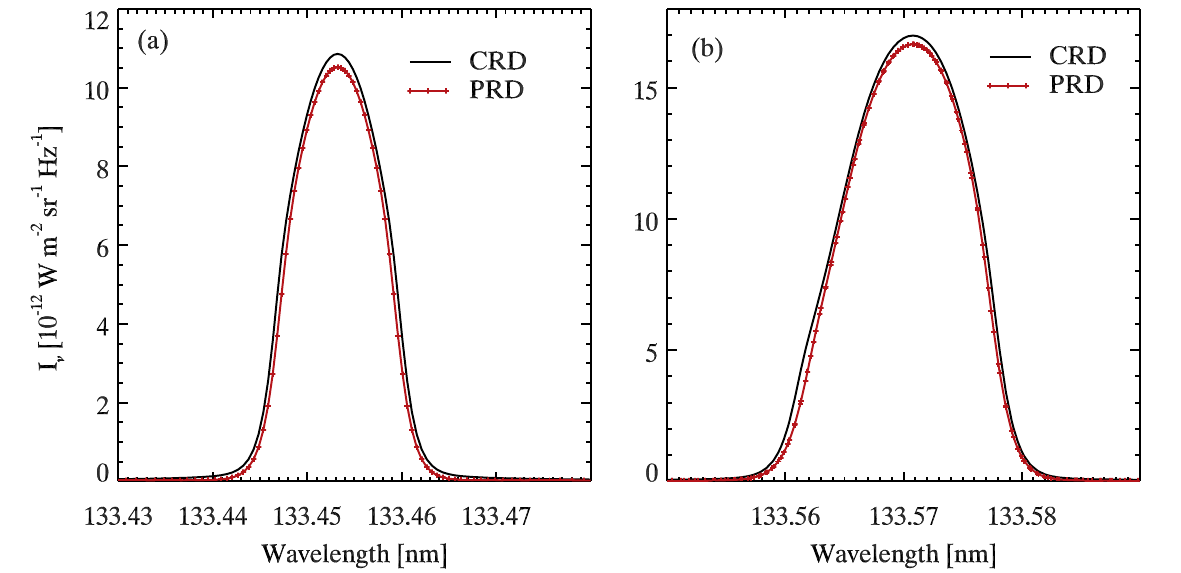}  
  \caption[]{\label{prd} 
Comparison between partial redistribution (PRD, red) and complete redistribution (CRD, black) approximation for the \cii\ lines in the VAL3C atmosphere for the \ciia\ line (a) and \ciib\ blend (b).}
\end{figure}

The \cii\ lines are resonance lines (lower levels part of the ground
term) and one might expect effects of partial frequency redistribution
of photons (PRD) to be important. We tested this by solving the
non-LTE problem for given atmospheres both including PRD effects and
using the computationally advantageous approximation of complete
redistribution (CRD). We did not find any significant differences, see
Figure~\ref{prd}, and conclude that CRD can be used in the modelling
of the \ciib\ lines under solar conditions.

%%%%%%%%%%%%%%%%%%%%%%%%%%%%%%%%%%%%%
%===========================================================================
%MB should have SI units for intensity with the units given in the axis label
%===========================================================================
%%%%%%%%%%%%%%%%%%%%%%%%%%%%%%%%%%%%%

\subsection {Importance of 3D radiative transfer}

\begin{figure}[hbtp]
  \includegraphics[width=\columnwidth]{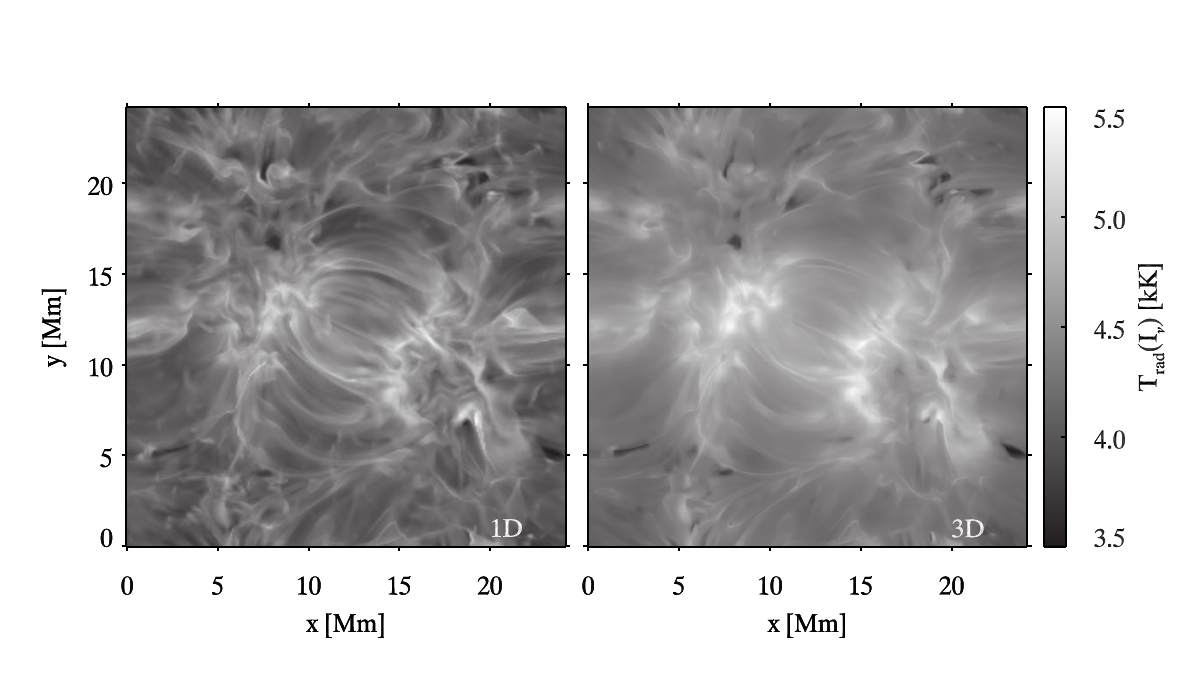}
  \caption[]{\label{fig:I3d}
  Comparison of the vertically emergent intensity in the core of the \ciia\ line solving each column in the 3D atmosphere 
  independently as a 1D problem (left) and  solving the full 3D radiative transfer problem (right).}
 \end{figure}

Figure~\ref{fig:I3d} shows a comparison of the vertically emergent intensity in the core of the \ciia\ line solving each column in the 3D atmosphere 
  independently as a 1D problem  and  solving the full 3D radiative transfer problem. It is clear that 3D scattering effects are important in the core of the line and a full 3D 
  transfer solution should be employed. It also means that we should expect the intensity close to the core to be influenced by the surroundings and not provide a very accurate mapping of the conditions along the local column. 
  
%%%%%%%%%%%%%%%%%%%%%%%%%%%%%%%%%%%%%
%===========================================================================
%===========================================================================
%%%%%%%%%%%%%%%%%%%%%%%%%%%%%%%%%%%%%

\section{Results}
\label{sec:res}

In this section we will study in detail how the \cii\ lines at 133.5$\,$ nm are formed. We use the quintessential 9-level model atom described above, assuming CRD but 
using the full 3D solution for the \bifrost\ atmosphere results.

\subsection{Contribution function to the intensity}\label{sec:cf}
% define contribution function
% define split for four panel diagrams
% detailed discussion of a few select cases
%   optically thick formation - maybe case with thin component
%   importance of source function - single peak vs double peak

As a tool to find out which parts of the atmosphere contribute to the emergent intensity of the \cii\ lines, we use the contribution function to intensity.

The emergent intensity in a 1D plane parallel semi-infinite atmosphere can be written as
%% Equation
%_______________________________________________________________________
\begin{equation}\label{emr_int}
I_{\nu,\mu}=\int_0^\infty \frac{1}{\mu} S_\nu (z) e^{-\tau_\nu (z)/ \mu} \chi_\nu dz
\end{equation}
%----------------------------------------------------------------------------------------------------------------------------------

In this equation, $\mu= \cos\theta$, where $\theta$ is the angle between the line of sight to the observer and the local vertical. The source function ($S_{\nu}$), opacity ($\chi_{\nu}$) and optical depth ($\tau_{\nu}$) are functions of frequency ($\nu$) and geometrical height ({\em{z}}). \edta{At disk center ($\mu=1$), 
%% Equation
%_______________________________________________________________________
%\begin{equation}\label{int}
%I_{\nu} =\int_0^\infty S_{\nu} e^{-\tau_{\nu}} \chi_{\nu} dz
%\end{equation}
%----------------------------------------------------------------------------------------------------------------------------------
the integrand in Equation~\ref{emr_int}} describes the local creation of photons ($S_\nu \chi_\nu dz$) and the fraction of those that escape ($e^{-\tau_\nu}$). It is thus natural to define the contribution function to intensity on a geometrical height scale as
%% Equation
%_______________________________________________________________________
\begin{equation}\label{eq:ci}
C_{I_{\nu}}(z)=S_{\nu} e^{-\tau_{\nu}} \chi_{\nu}.
\end{equation}
%----------------------------------------------------------------------------------------------------------------------------------

Following \citet{1997ApJ...481..500C} we also rewrite the contribution function as

\begin{equation}\label{eq:cisplit}
C_{I_{\nu}}(z)=S_{\nu} \tau_\nu e^{-\tau_{\nu}} {\chi_{\nu} \over \tau_\nu}.
\end{equation}

where the term $\tau_\nu e^{-\tau_{\nu}}$ having a maximum at $\tau_\nu=1$ represents the Eddington-Barbier
part of the contribution function, $S_\nu$ gives the source function contribution and the final term,
${\chi_{\nu} \over \tau_\nu}$ picks out effects of velocity gradients in the atmosphere.

To illustrate the range of line formation scenarios we have picked four different columns from the \bifrost\
atmosphere. These columns are marked in Figure~\ref{int_mag}.

\begin{figure}[hbtp]
 \includegraphics[width=\columnwidth]{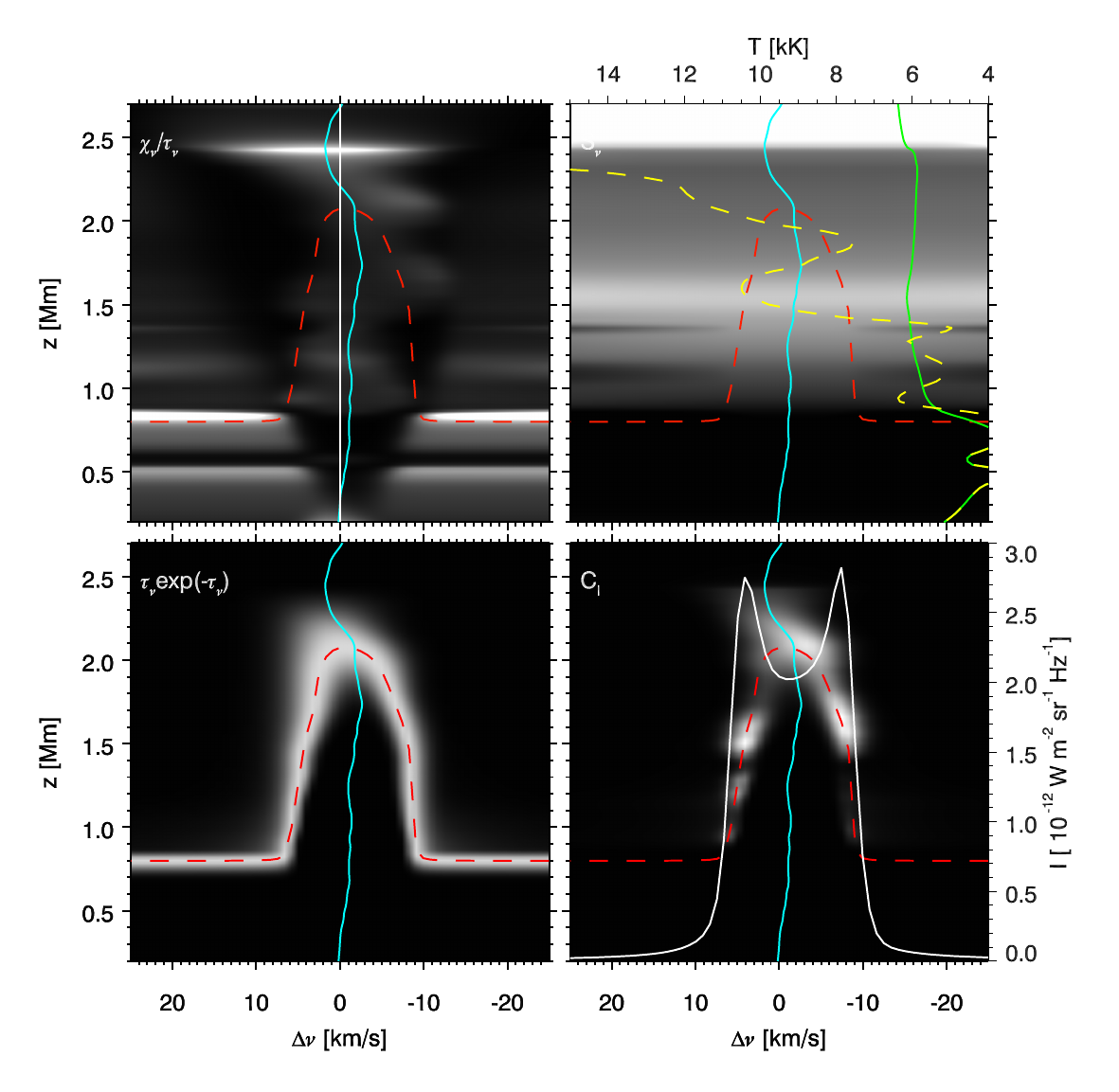} 
  \caption[]{\label{fig:cia} 
   Contribution function to intensity of the \ciia\ line.
   Each panel shows the quantity specified in
    the top-left corner as a grey-scale image as function of frequency from line center (in
    Doppler shift units) and atmospheric height $z$, see equation~\ref{eq:cisplit}.  Multiplication of
    the first three produces the contribution function to intensity shown in
    the lower right panel. A $\tau_\nu\is1$ curve (red dashed) and the
    vertical velocity (turquoise solid, positive is upflow) are
    shown in each panel, with a $v_z \is 0$ line in the upper left
    panel for reference.  The upper-right panel also contains the
    Planck function (yellow dashed) and the total source function (green solid) in radiation temperature 
    units specified along the
    top. The lower-right panel also contains the emergent intensity
    profile (white solid), with the scale along the
    right-hand side.
    The formation diagram is shown for  column $(x,y)=(12,10.3)\,$Mm in the \bifrost\ atmosphere (column (A) in Figure~\ref{int_mag}).
   }
\end{figure}

\begin{figure}[hbtp]
  \includegraphics[width=\columnwidth]{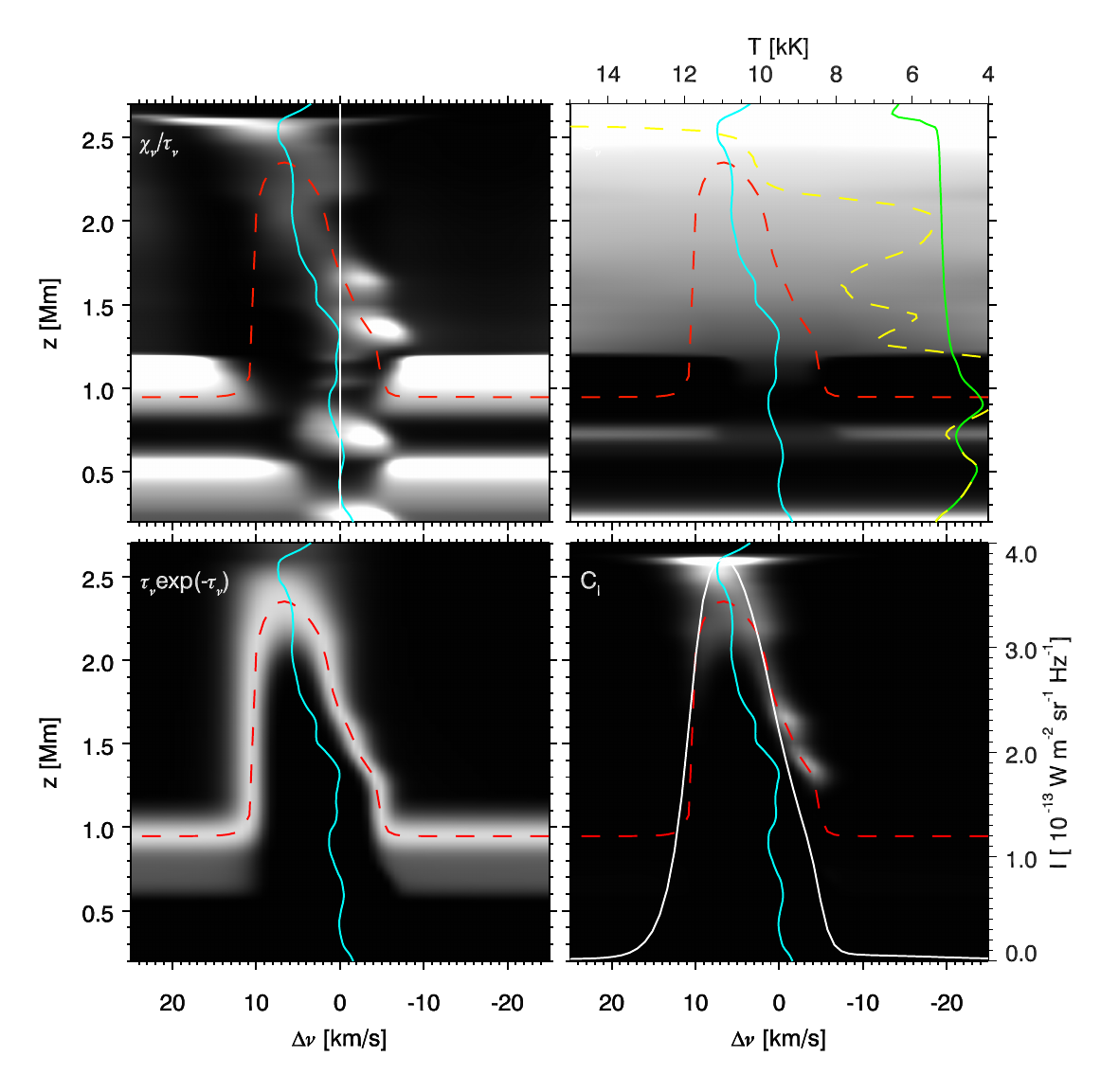} 
  \caption[]{\label{fig:cib} 
   As Figure~\ref{fig:cia} but for  column $(x,y)=(12,22.8)\,$Mm in the \bifrost\ atmosphere (column (B) in Figure~\ref{int_mag}).
   }
\end{figure}

\begin{figure}[hbtp]
  \includegraphics[width=\columnwidth]{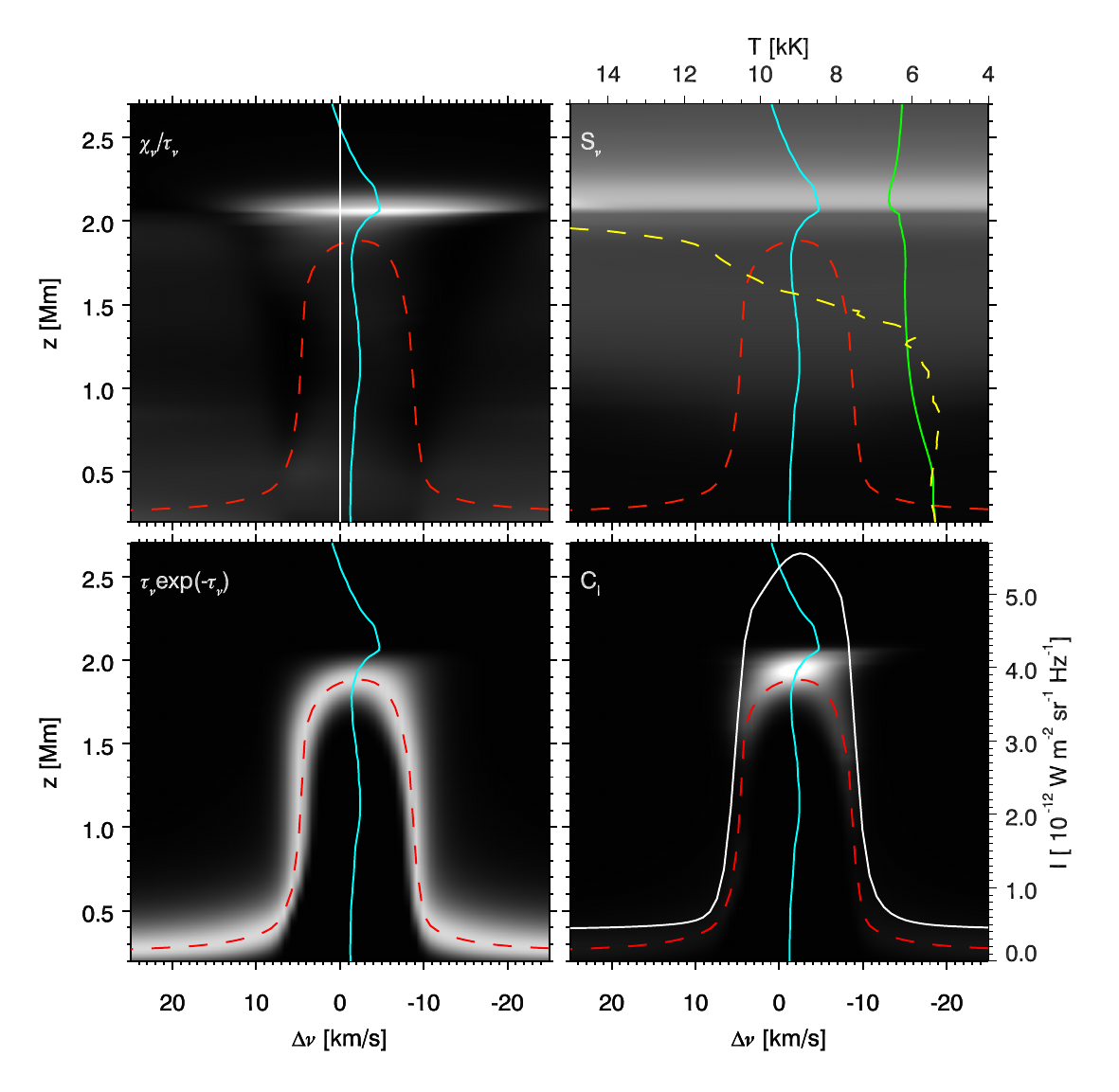} 
  \caption[]{\label{fig:cic} 
   As Figure~\ref{fig:cia} but for  column $(x,y)=(6.9,12.9)\,$Mm in the \bifrost\ atmosphere (column (C) in Figure~\ref{int_mag}).
   }
\end{figure}

\begin{figure}[hbtp]
  \includegraphics[width=\columnwidth]{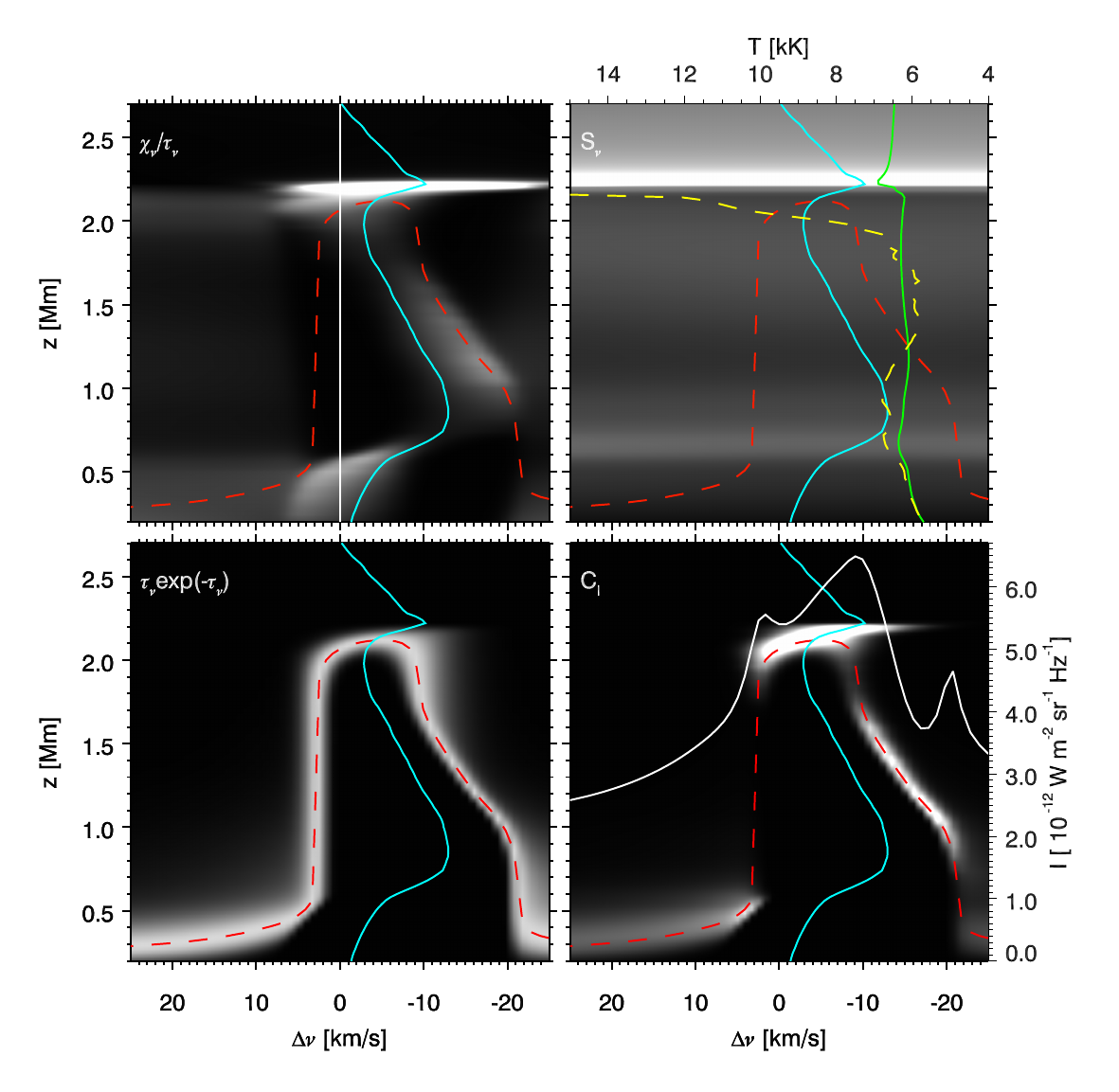} 
  \caption[]{\label{fig:cid} 
   As Figure~\ref{fig:cia} but for  column $(x,y)=(14.5,9.9)\,$Mm in the \bifrost\ atmosphere (column (D) in Figure~\ref{int_mag}).
   }
\end{figure}

Figures~\ref{fig:cia}--\ref{fig:cid} show the total contribution function as function of height and
frequency across the \ciia\ line (bottom right panels) as well as the three individual terms above in
the same manner as in the four-panel diagrams in \citet{1997ApJ...481..500C} for these four columns. To avoid the confusion from the blend we have chosen the $133.4\,$nm line to analyze the
formation in detail but the same general characteristics are true for the stronger line as well but with a
formation a bit higher in the atmosphere.

%MM we start the discussion with the most typical profile: doubly peaked
%MM detailed discussion of the formation - why do we have emission, where is the contribution function
%MM peaked, importance of velocity gradient, etc
%MM for each column we should describe what kind of column this is: internetwork, 
%MM magnetic field concentration, network etc
Figure~\ref{fig:cia} shows a typical case of a double peaked profile (the intensity profile is shown
in the bottom right panel). The velocity (turquoise  almost vertical line in all panels) does not exhibit
strong gradients which is why the $\chi_\nu/\tau_\nu$ term in the upper left panel does not exhibit
strong variations with wavelength or height. The $\tau_\nu\is 1$ (red dashed line in all panels) ranges from 0.8~Mm in the
continuum to 2.1~Mm at line center. The source function (solid green line in the upper right panel)
is equal to the Planck function (the dashed yellow line shows the radiation temperature of the
Planck function and thus the temperature itself) from the photosphere up to a height of 0.7~Mm where
it decouples. The strong peak in temperature at a height of 1.6~Mm causes a peak in the source
function (not so pronounced in the radiation temperature representation shown with the green line
but quite evident in the linear representation shown in the image in the upper right panel). This
peak in the source function is responsible for the two emission peaks in the intensity profile.
Because of the lack of strong velocity gradients, the intensity profile is quite symmetric with
almost equal intensity of the read and blue peaks. The line core is formed at the highest height of the
optical depth unity curve at 2.1~Mm. The source function is there lower than at 1.6~Mm and we get a
central reversal. At 2.1~Mm height there is a downward velocity of 2 km~s$^{-1}$ which causes a
redshift of the line center with a similar amount.

%MM optically thin, singly peaked profile
Figure~\ref{fig:cib} shows a case of a single peaked profile where the formation is optically thin.
It is clear from the total contribution function image (lower right panel) that the maximum of the
contribution to intensity comes from a height of around 2.6~Mm which is well above the maximum
\edt{height of optical depth unity} (2.3~Mm). The atmosphere is moving upwards with 7 km~s$^{-1}$ where the
intensity is formed and this gives a blueshift of the same amount. Note that the whole profile has a
 blueshift of 5-7 km~s$^{-1}$ even though the atmospheric velocity is
almost zero below a height of 1.3~Mm. This is because most of the intensity profile is formed
around 2.6~Mm height and not at the optical depth unity height. The asymmetry with a stronger red
wing than blue wing comes from the velocity gradient giving a larger $\chi_\nu/\tau_\nu$
term (upper left panel) on the red side. The full-width-half-maximum (FWHM)
of the intensity profile is 10.6 km~s$^{-1}$ corresponding to the thermal broadening at 29 kK which is
consistent with the temperature in the lineforming region.
% steep gradient. T[z=2.600]=24kK, T[z=2.604]=29 kK, T[z=2.61]=38 kK

%MM magnetic concentration
Figure~\ref{fig:cic} shows the formation of a single peaked profile in a flux concentration (column C in 
Figure~\ref{int_mag} at $(x,y)=(6.9,12.9)\,$Mm). Because of the strong magnetic field (-1.6 kG in the photosphere), the
density is lower than in the surroundings and a given optical depth occurs at a lower geometrical height. The continuum
is therefore formed at 0.25~Mm compared with the average continuum formation height of 0.8~Mm. The core of the line
is also formed slightly lower than in the surroundings, at 1.9~Mm height, but close to the height where the monochromatic
optical depth is unity. The line thus shows an optically thick formation with the intensity being a map of the source function at 
optical depth unity. The source function is monotonically increasing with height up to where the core is formed so there
is no central depression. Note also the rather box-shaped profile - a consequence of a source function that increases more
rapidly just above the continuum formation height than further up towards the core formation height.

%MM next magnetic concentration
Finally, Figure~\ref{fig:cid} shows the formation of the intensity in another flux concentration (column D in 
Figure~\ref{int_mag} at $(x,y)=(14.5, 9.9)\,$Mm). The continuum is formed low because of the strong magnetic field 
(1.7 kG in the photosphere). 
The continuum intensity is high because of the high temperature at optical depth unity in the continuum (see also
Figure~\ref{fig:atmos_mod}).
Above the photosphere we have strong velocity gradients that cause a broadened asymmetric profile.
The source function is increasing up to a height of 0.6~Mm, is decreasing up to the height of optical depth unity 
at line center  and then shows 
a sharp increase.We therefore get a profile with three peaks with the outer ones formed at the wavelengths 
where optical depth unity is at 0.6~Mm. The FWHM of the intensity profile is 26.1 km~s$^{-1}$ which
is much larger than the thermal width in the line forming region. This is partly because of the large velocity
gradients in the atmosphere at this column but also the profile 
calculated without velocity gradients in the atmosphere has a profile much broader than what would be
expected from the thermal width. This
is typical for a line that is formed in the optically thick regime. The line-width depends strongly on how the source function varies with 
height in the atmosphere. We will return to this point in Section~\ref{sec:width}

%%%%%%%%%%%%%%%%%%%%%%%%%%%%%%%%%%%%%
%===========================================================================
%===========================================================================
%%%%%%%%%%%%%%%%%%%%%%%%%%%%%%%%%%%%%

%%%%%%%%%%%%%%%%%%%%%%%%%%%%%%%%%%%%%
%===========================================================================
%===========================================================================
%%%%%%%%%%%%%%%%%%%%%%%%%%%%%%%%%%%%%

%%%%%%%%%%%%%%%%%%%%%%%%%%%%%%%%%%%%%
%magnetic footprint Dark on the left
%===========================================================================
%===========================================================================
%%%%%%%%%%%%%%%%%%%%%%%%%%%%%%%%%%%%%

%%%%%%%%%%%%%%%%%%%%%%%%%%%%%%%%%%%%%
%magnetic footprint bright on the right
%===========================================================================
%===========================================================================
%%%%%%%%%%%%%%%%%%%%%%%%%%%%%%%%%%%%%

% line center formation

So far we have exemplified the formation of the \ciia\ line through a detailed analysis of four different
columns in the \bifrost\ atmosphere. We will now look at the formation of the intensity along the cut at $x\is 12\,$Mm 
(see Figure~\ref{int_mag}). We will investigate the contribution function to intensity at the line core; where
for each column in the simulation atmosphere, the line core is defined as the wavelength with maximum intensity (single peaked profiles) or minimum 
intensity in the central reversal (double peaked profiles). This contribution function is given in Equation~\ref{eq:ci}.
We are also interested in where the integrated intensity is coming from. We achieve this by defining the
contribution function to total intensity as the integral over frequency of the contribution function to 
specific intensity:

\begin{equation}\label{ci_tot}
 C_{I_{\rm total}} (z)= \int C_{I_{\nu}}(z) \ d\nu 
\end{equation}

Note that this definition only makes sense when the continuum intensity is much lower than the line intensity
(which is the case here). We perform the integration over a wide enough frequency range to cover the 
width of the line but narrow enough not 
to be dominated by the continuum intensity contribution.

\begin{figure}[hbtp]
  \includegraphics[width=\columnwidth]{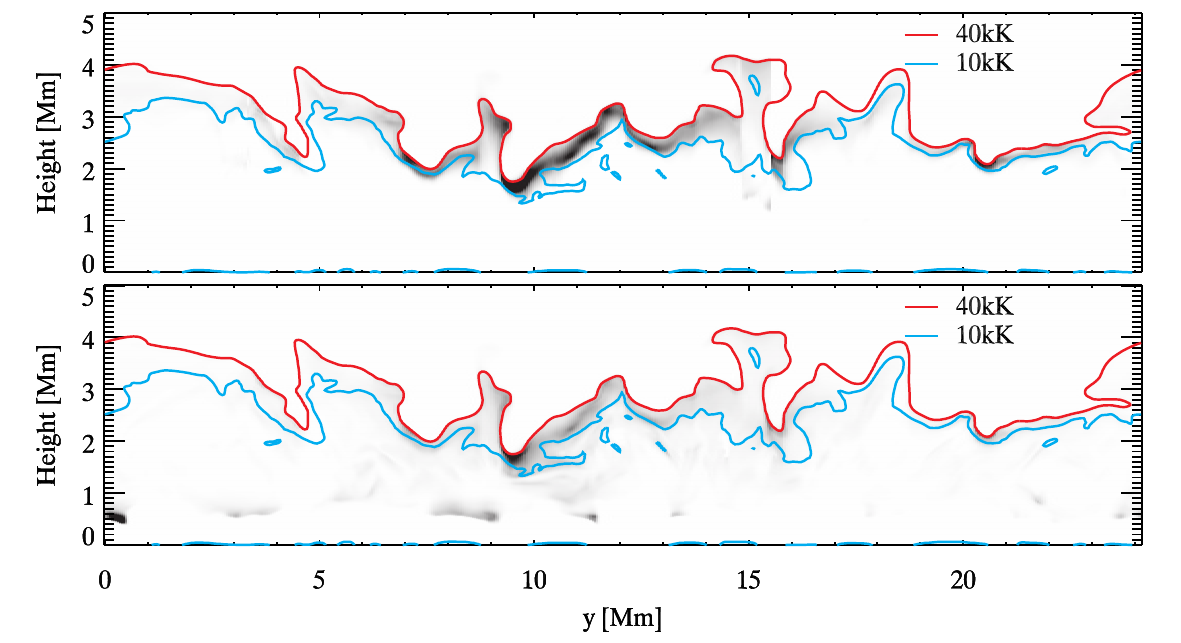}
  \caption[]{\label{cntb_cut} 
Contribution function to intensity at line core (top) and the contribution function to
total intensity (bottom) on a height scale along the cut through the \bifrost\ atmosphere 
at X\is 12~Mm shown as grey-scale images. Temperature contours are given at 40 kK (red) and
10 kK (blue). The integration for the contribution function to total intensity was performed 
over $\pm 20$ km~s$^{-1}$
}
 \end{figure}

The top panel of Figure~\ref{cntb_cut} shows the contribution function to the intensity at the line core  along this cut.
% here follows discussion of line center formation, comparison with temperature etc
The line core is normally formed between 10 kK and 40 kK but there are also regions \edta{where the 
line core contribution function reaches its maximum at lower temperatures} (e.g. around $y\is17-18\,$Mm). 
This happens when the transition region is at a low 
density. 
Optical depth unity is \edta{reached when we have high enough density. With a low density in the transition region, this
happens deeper down where the temperature is lower.}
Because of the lower \edta{temperature}, the line is also weaker 
(since the intensity is the integral of the contribution function,
Figure~\ref{cntb_cut} also gives an indication of the intensity). The formation layer of the line core is 
between 2 and 4 Mm height and is as corrugated as the
temperature structure.

The bottom panel of Figure~\ref{cntb_cut} shows the contribution function to total intensity. 
The total intensity is mostly coming
from the core formation region but there is also a broad tail of the contribution function 
going all the way down to the formation
region of the continuum (which is on average around a height of 0.8 Mm). 
At $y\is0-0.5\,$Mm the contribution function to total
intensity is dominated by the continuum contribution. Here the continuum is enhanced because of a 
high temperature(up to 7~kK) around 0.8~Mm and the line is actually in absorption.

\edta{The height of unit optical depth} (and therefore also the formation height) varies with the density stratification.
Since the $133.5\,$nm lines are resonance lines, we expect the optical depth unity to be at given column density
of  singly ionized carbon. The opacity at the line core depends on the broadening of the profile but is 
on the order of $10^6\,{\rm cm}^2\, {\rm g}^{-1}$ for the \ciia\ line. We therefore expect optical depth unity to be at a column mass 
of about $10^{-6}\, {\rm g\, cm}^{-2}$. We can define a contribution function on a logarithmic column 
mass scale from
\begin{equation}
C_{I_\nu}(\log_{10} m_c) d(\log_{10} m_c)=C_{I_\nu}(z)dz
\end{equation}
where $m_c$ is the column mass, which leads to
\begin{equation}\label{eq:cim}
C_{I_\nu}(\log_{10} m_c) ={m_c\over\rho}\ln(10)C_{I_\nu}(z).
\end{equation}

\begin{figure}[hbtp]
  \includegraphics[width=\columnwidth]{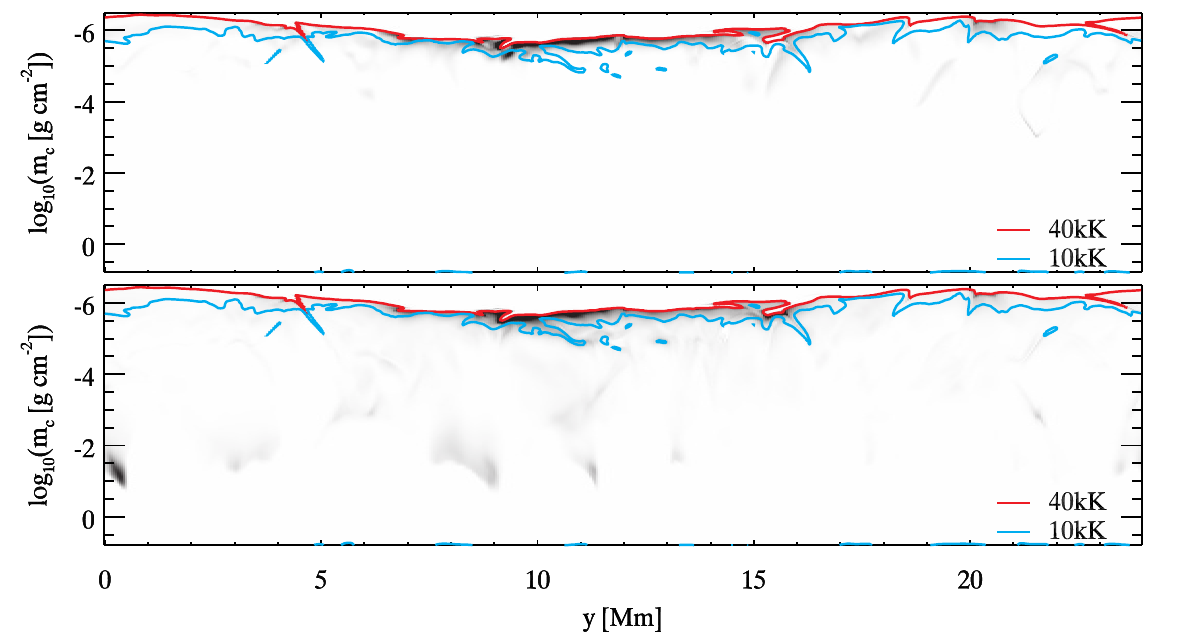}
  \caption[]{\label{cntb_log_cm_cut} 
As Figure~\ref{cntb_cut} but on a 
logarithmic column mass scale (see Equation~\ref{eq:cim})
}
 \end{figure}

Figure~\ref{cntb_log_cm_cut} shows the contribution function to intensity at line core 
(top panel) as well as the contribution function to total
intensity (bottom panel) on a logarithmic column mass scale. The range in logarithmic column
mass has been chosen to correspond to the range in height for y=0\edta{. This figure illustrates} clearly that the
%this may seem contradicted by the fact that the T=40kK curve is at 4Mm at y=0 but close to the top in 
%the column mass scale. This is because the column mass does not change much with height in the range 4-8 Mm. 
%At 4Mm lg(mc)=-6.38, at 5Mm lg(mc)=-6.47, at 6 Mm lg(mc)=-6.54
corrugation of the transition region is much smaller on a column mass scale than on a height scale
and that the contribution to intensity, both at line core and for the integrated total intensity, is
concentrated to a small range in column mass. We can also see that indeed the contribution
to intensity is around a column mass of $10^{-6}\, {\rm g\, cm}^{-2}$.

In sum we expect line core diagnostics (especially the shift of the core) to give 
information from the region just below the transition
region in the quiet Sun. Diagnostics using the full line (like moments of the intensity) will 
be influenced by a much larger part of the chromosphere. We will return to the statistical correlation between
various line shift diagnostics and the physical velocity in the simulation in Paper II.

%===========================================================================
%%%%%%%%%%%%%%%%%%%%%%%%%%%%%%%%%%%%%
%===========================================================================
%%%%%%%%%%%%%%%%%%%%%%%%%%%%%%%%%%%%%

\subsection{Line width}\label{sec:width}

For an optically thin line \edta{formed in the CRD regime}, the width of the intensity profile is directly given by the width of the absorption/emission profile 
\begin{equation}\label{eq:phi}
\Phi_\nu={1\over\sqrt{\pi}\Delta\nu_D}e^{-({\Delta\nu\over\Delta\nu_D})^2}
\end{equation}
where $\Delta\nu$ is the difference in frequency from the line centre frequency, $\nu_0$, and
\begin{equation}
\Delta\nu_D={\nu_0\over c}\sqrt{{2 k T\over m}+\xi^2}
\end{equation}
where $\sqrt{2 k T/m}$ gives the thermal Doppler broadening and $\xi$ is a non-thermal velocity.
Here $k$ is the Boltzmann constant, $m$ is the mass of the atom/ion and we
have assumed a Maxwellian velocity distribution at the temperature $T$. In 
an isothermal plasma and in the absence of a macroscopic
velocity field we get a symmetric Gaussian intensity profile with maximum
intensity at $\nu_0$ and a full-width of $2\Delta\nu_D$ at $1/e$ of the maximum intensity. Other popular
measures of line width include the standard deviation, $\sigma$, of a Gaussian fit
\begin{equation}
I_\nu=(I_0-I_c) e^{-{\Delta\nu^2\over 2\sigma^2}}+I_c,
\end{equation}
the full-width at half the maximum intensity ($W_{\rm FWHM}$), and the second moment of intensity
\begin{equation}
W_2=\sqrt{\int{(\nu-\nu_1)^2 (I_\nu-I_c) d\nu}/I_{\rm tot}}
\end{equation}
where
\begin{equation}
I_{\rm tot}=\int{(I_\nu-I_c)d\nu}
\end{equation}
is the zeroth moment and
\begin{equation}
\nu_1=\int{\Delta\nu(I_\nu-I_c)d\nu}/I_{\rm tot}
\end{equation}
is the first moment.

For a Gaussian intensity profile we have the following relations between the four width measures:
\begin{eqnarray}
\Delta\nu_D&=&\sqrt{2}\sigma\\
W_{\rm FWHM}&=&2\sqrt{\ln 2}\Delta\nu_D\\
W_2&=&\sigma.
\end{eqnarray}

\begin{figure}[hbtp]
  \includegraphics[width=\columnwidth]{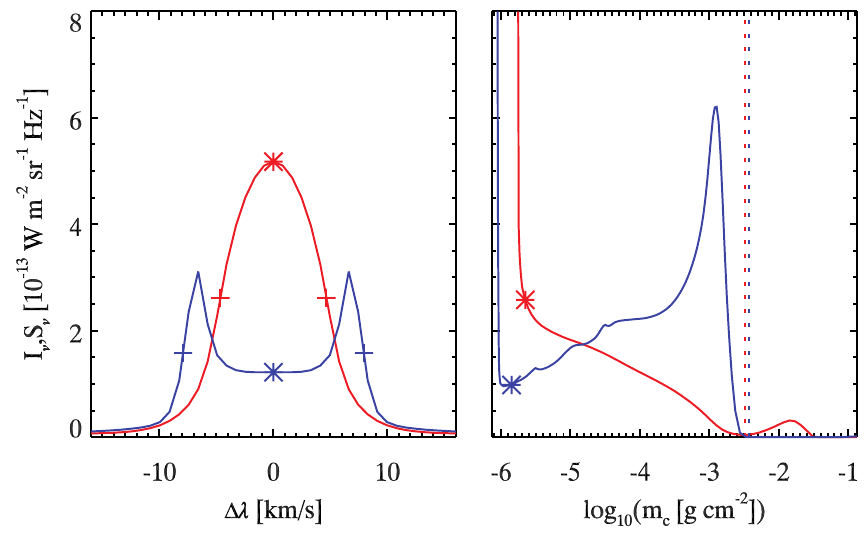}
  \caption[]{\label{width_snum} 
  Intensity as function of wavelength (given as equivalent velocity) at two columns in the \bifrost\ atmosphere with 
  zero velocities ({left}) and corresponding source functions as function of logarithmic column 
  mass ({right}). The intensities at the line core are marked with asterisks in the left panel and the 
  corresponding height for optical depth unity in the right panel. The heights for optical depth unity in the
  continuum are marked with vertical dotted lines. The points corresponding to the FWHM width are marked
  with plus-signs in the left panel with widths of 9.3 km/s ({red}) and 15.9 km/s ({blue}).
  The two positions are at $(x,y)\is(12,7.5)\,$Mm (red) and
  $(x,y)\is(12,6.1)\,$Mm (blue).

}
 \end{figure}

For an optically thick line (as our \cii~lines) the situation is different. The width of the absorption profile 
still plays a role but in addition the source function variation with depth is crucial. To illustrate this we
have carried out a full 3D non-LTE calculation with the \bifrost\ atmosphere described \edt{earlier,} but with all
velocities set to zero. We thus have no effects from non-thermal velocities and the \ciia\ intensity profile
is symmetric. In Figure~\ref{width_snum} we show the intensity profiles and corresponding source functions
for two columns in this simulation. The red profile is illustrating a narrow, single-peaked profile where the 
source function increases almost linearly with decreasing logarithmic column mass from the point in the atmosphere where optical depth is unity
in the continuum ($\log_{10}m_c=-2.5$) to optical depth unity in the core ($\log_{10}m_c=-5.6$). The 
temperature in the line-forming region is 11 kK which gives a
thermal broadening ($\sqrt{2 k T/m}$) of 4 km/s and a thermal $W_{\rm FWHM}=6.7$~km/s.
For the red profile we find $W_{\rm FWHM}=9.3$~km/s which is a factor 1.4 larger than the thermal 
width, even in this model where there are no non-thermal motions. This is sometimes called
opacity broadening of an optically thick line. This broadening is dependent on the source function 
variation with depth. If the source function has a peak in the low atmosphere, just above the formation height of the continuum, 
we get steep emission flanks, a double peak intensity profile and a larger width. The blue profile in
Figure~\ref{width_snum} is an
example of such a scenario with $W_{\rm FWHM}=15.9$~km/s with a temperature in the formation region of 9.6 kK.

%===========================================================================
%%%%%%%%%%%%%%%%%%%%%%%%%%%%%%%%%%%%%

We can get an estimate of the "opacity broadening" factor by using the Eddington-Barbier relation
(that states that the emergent vertical intensity is equal to the source function at optical depth unity).

The optical depth is given by
\begin{equation}
d\tau_\nu=-\chi_\nu dz
\end{equation}
where $\chi_\nu$ is the opacity per volume. The column mass is given by
\begin{equation}
dm_c=-\rho dz
\end{equation}
with $\rho$ being the mass density. We thus have
\begin{equation}
d\tau_\nu={\chi_\nu\over\rho} dm_c
\end{equation}
and we get the column mass where $\tau_\nu=1$ at frequency $\Delta\nu$ from
\begin{equation}
1=\int_0^{m_c(\Delta\nu)}{\chi_\nu\over\rho}dm_c
\end{equation}
In an isothermal atmosphere with no non-thermal velocities we have a constant atomic absorption profile
and for a resonance line of the dominant ionization stage we get a constant line opacity per unit mass:
\begin{equation}
{\chi_{l\nu}\over\rho}=\alpha_l e^{-({\Delta\nu\over\Delta\nu_D})^2}
\end{equation}
with $\alpha_l$ a constant. 
We assume
the same to be true also for the continuum opacity, $\chi_c$ (where $\chi_\nu=\chi_{l\nu}+\chi_c$). We then get

\begin{equation}
{1\over m_c(\Delta\nu)} = \alpha_l e^{-({\Delta\nu\over\Delta\nu_D})^2}+{\chi_c\over\rho}
\end{equation}
Using the same equation for the continuum, $\Delta\nu\rightarrow\infty$, we can eliminate
the $\chi_c/\rho$ term:
\begin{equation}
{1\over m_c(\Delta\nu)} - {1\over m_c(\infty)}= \alpha_l e^{-({\Delta\nu\over\Delta\nu_D})^2}
\end{equation}
For the line centre, $\Delta\nu=0$, we get
\begin{equation}
{1\over m_c(0)} - {1\over m_c(\infty)}= \alpha_l 
\end{equation}
This gives
\begin{equation}\label{eq:taucm}
{\Delta\nu\over\Delta\nu_D}=\sqrt{\ln{{1\over m_c(0)} - {1\over m_c(\infty)}\over{1\over m_c(\Delta\nu)} - {1\over m_c(\infty)}}}.
\end{equation}
In the case $m_c(\infty) \gg m_c(\Delta\nu)$ this reduces to
\begin{equation}\label{eq:taucm2}
{\Delta\nu\over\Delta\nu_D}=\sqrt{\ln{m_c(\Delta\nu)\over m_c(0)}}.
\end{equation}

% blue profile
% lgmc0=-5.8475 & lgmc1=-2.89571 & lgmcc=-2.4321 ; for peaks where v=7.02 km/s => 
% wopc=2.69, w_th=6.1, T=9640
% red profile
% lgmc0=-5.649 & lgmc1=-4.067 & lgmcc=-2.484 => wopc=1.9
For  the blue profile in Figure~\ref{width_snum} we have for the FWHM formation height
$\log_{10}(m_c(\Delta\nu))\approx -2.7$, $\log_{10}(m_c(0))= -5.8$ and 
$\log_{10}(m_c(\infty))= -2.4$. \edta{Using Equation~\ref{eq:taucm},} this gives an opacity 
broadening factor of 2.8, which is similar to the
actual value of 15.9/6.1=2.6. 
For the red profile we get an
opacity broadening factor of 1.9, which is larger than the actual value of 1.4. This is because
the Eddington-Barbier relation underestimates the intensity peak when the source function
is very steep and non-linear in the line forming region and therefore overestimates the value of $m_c(\Delta\nu)$
\edta{that is necessary to reach optical depth unity}.
Equation~\ref{eq:taucm} also
gives the correct position of the emission peaks of the blue profile. 
We conclude that the width of the \cii~lines can be
1.4--3 times what is given by the width of the atomic absorption profile with larger values for
double peaked profiles.

\subsection{Intensity ratio}\label{sec:iratio}

The intensity ratio between the components in the multiplet gives additional diagnostic value. 
Since the two levels within each term are very close in energy, collisions will be efficient in driving
the population ratio towards the ratio of the degeneracies. The lower levels have a ratio of their
degeneracies of two and our simulations confirm that the population densities have this ratio
throughout the line forming regions. The opacity scales as $n_l f_{lu}$ where $n_l$ is the
population density of the lower level and $f_{lu}$ is the absorption oscillator strength. The oscillator
strengths of the two lines have a ratio of 0.9 such that the opacity in the $133.5708\,{\rm  nm}$ line is 
a factor of 1.8 times that of the \ciia\ line. 
In the optically thin limit, we have
\begin{equation}
I_\nu\simeq\int(\chi_{l\nu}+\chi_c)S_\nu ds
\end{equation}
In the coronal approximation (collisional excitation, radiative deexcitation), the source functions of the
two lines are the same and the line opacity, $\chi_{l\nu}$, has the above ratio of 1.8 because the atomic absorption
profiles are also identical. The intensity ratio is thus 1.8 at all frequencies across
the line profiles where the line opacity dominates over the \edta{continuum} opacity. 

For an optically thick line, this is no longer true. We start by discussing the case of double peaked profiles and
equal source functions.
The two emission peaks of each line are formed at the local source function maximum and we therefore 
get equal peak intensities (but with a larger separation for the stronger \ciib\ line). 
The self reversal is formed higher up where the source function is decreasing with height
and the stronger line has its core formed higher and therefore has a lower core intensity.
Single peak profiles mean that the source function is increasing with height and the \ciib\ line will therefore have
higher peak intensity than the \ciia\ line. In general, the two source functions are {\em not} equal and we can in 
principle get any intensity ratio. The weakest blend share upper level with the \ciia\ line and since it becomes 
optically thin deeper down than the other lines, it tends to depopulate the upper level decreasing the source function
also of the \ciia\ line. This source function difference tends to increase the intensity ratio between the lines.

\begin{figure}[hbtp]
  \includegraphics[width=\columnwidth]{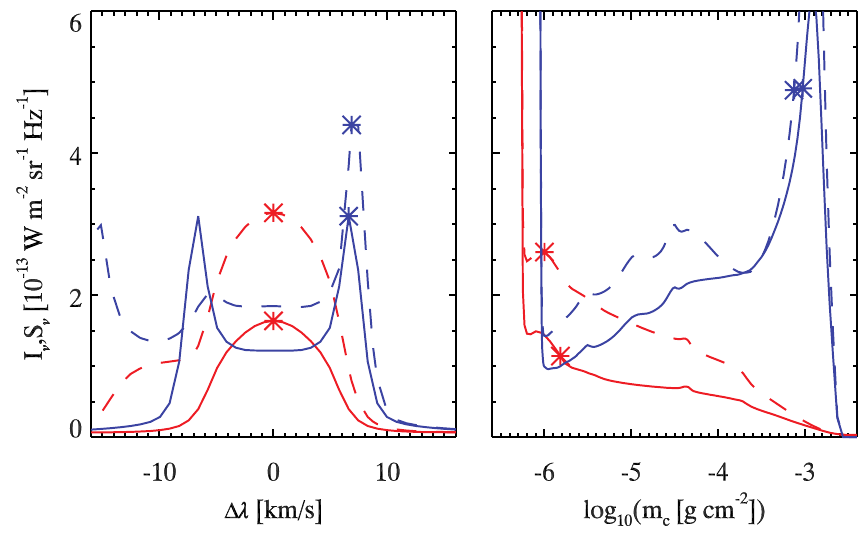}
  \caption[]{\label{ratio_snum} 
  Intensity as function of wavelength (given as equivalent velocity) at two columns in the \bifrost\ atmosphere with 
  zero velocities ({left}) for the \ciia\ line (solid) and the \ciib\ line (dashed) and corresponding source functions as function of logarithmic column 
  mass ({right}). Maximum intensities are marked with asterisks in the left panel and the 
  corresponding height for optical depth unity in the right panel. The two \bifrost\ columns are at $(x,y)\is(12,21)\,$Mm (red) and
  $(x,y)\is(12,6.1)\,$Mm (blue).
}
 \end{figure}

We illustrate the dependency of the ratio on the source function behaviour in Figure~\ref{ratio_snum} where we 
show intensity profiles and source functions for the two lines at two locations in the \bifrost\  simulation. 
The velocities were set to zero before the Multi3D calculation of intensities. We thus get symmetric intensity 
profiles for the \ciia\ line but asymmetric profiles for the \ciib\ line because of the blend in the blue wing. The
profiles in red color were taken from the column at $(x,y)=(12,21)\,$Mm where we have single peak intensity profiles (left panel).
The locations of line core optical depth unity are marked on the source functions displayed in the right panel. The
\ciib\ line core (dashed line) is formed higher than the \ciia\ line (lower column mass) and that would by itself give a higher 
core intensity since the source function is increasing with decreasing column mass. The main reason for the 
high intensity ratio of 1.9 is the large difference in the source functions. The profiles in blue color were taken from the
column at $(x,y)=(12,6.1)\,$Mm (same as the blue colored profile in Figure~\ref{width_snum}) where the intensity profile has
two peaks. For the stronger \ciib\ line we do not reach low enough opacity on the blue (short wavelength) side of the stronger component 
to reach optical depth unity at the source function maximum depth and we need to go all the way to the blue side of the
blend to get the corresponding intensity peak. This explains the rather weird profile (dashed blue line in the left panel).
The red (long wavelength) peaks have an intensity ratio of 1.4, again mainly caused by differences in the source functions.

The source function differences are driven by radiative rates, both losses in the weakest line decreasing the source 
function of the \ciia\ line that shares the upper level with the weak blend and absorptions of these photons in the
\ciib\ line (which lead to an increase of the \ciib\ source function). Collisions between the upper $2p\,^2\!D$ levels 
are not efficient enough to drive the source functions to equality. We have also included collisions with neutral hydrogen
between these levels but a factor of 10 higher total collisional rates would be needed to significantly affect the source function
ratios.

We conclude that the intensity ratio of the \ciib\ and \ciia\ lines is primarily indicating the ratio of the source functions of the lines. 
This ratio is normally larger than one but can have any value, including the optically thin value of 1.8. We expect a lower 
ratio when densities are high in the line forming region, as is the case for the intensity peaks of double peak profiles.

%===========================================================================
%%%%%%%%%%%%%%%%%%%%%%%%%%%%%%%%%%%%%

\subsection {Comparison with optically thin modelling}

\begin{figure}[hbtp]
  \includegraphics[width=\columnwidth]{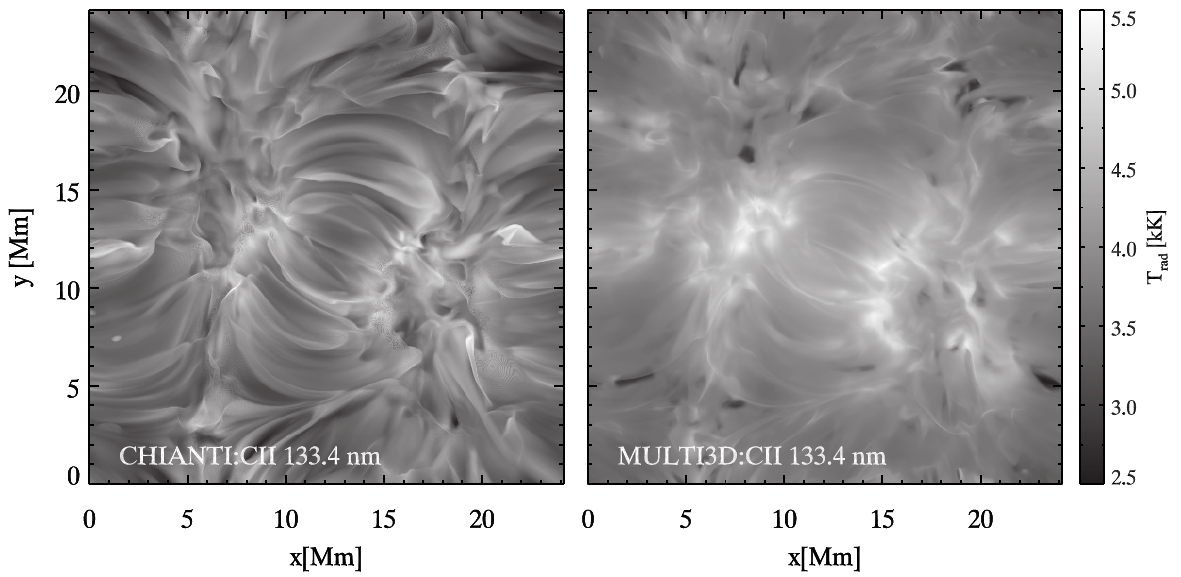} 
  \caption[]{\label{mulch_img} 
 Emergent intensity at 133.4663 nm from the coronal approximation (left panel) and from the full
 3D non-LTE calculation (right panel).  }
 \end{figure}

We have found that the \cii\ lines are mostly formed in the optically thick regime. If we instead assume optically 
thin formation with the ionization equilibrium set by collisional ionization and dielectric recombination without
density dependent effects (standard coronal approximation) we severely underestimate the amount
of \cii\ in the upper chromosphere and overestimate the amount
in the higher temperature region (see Figure~\ref{mulch_ioneq}). In the optically thin regime, the intensity is set by the 
source function multiplied by the optical thickness of \cii\ in the region. With the low densities where \cii\ exists
in the coronal approximation, this gives an optical thickness of less than 0.1. The source function then needs to be
more than ten times higher than what it is in our model at optical depth unity to give a similar intensity. 
This is normally not the case and we thus
get a weaker line from the coronal approximation than in our non-LTE modelling. This is shown in Figure~\ref{mulch_img}
where we compare the emergent intensity from assuming the coronal approximation (left panel) with our simulation employing the full 3D non-LTE modelling (right panel).

%%%%%%%%%%%%%%%%%%%%%%%%%%%%%%%%%%%%%  
%===========================================================================
%===========================================================================
%%%%%%%%%%%%%%%%%%%%%%%%%%%%%%%%%%%%%

\section{Sunspot Model}\label{sec:sunspot}

In addition to the quiet sun case, we have also studied the behavior of the lines in the sunspot 
model atmosphere of  \citet{1982ApJS...49..293L}.  The temperature structure of this model is shown
in Figure~\ref{fig:atmos_mod}. 
The photospheric temperature is naturally lower than in the quiet Sun model VAL3C and a
low temperature of 3.5~kK continues up to $\log_{10}m_c\is -2$ where there is a rapid temperature
increase to a chromospheric plateau of 6.5~kK, slightly above the VAL3C temperatures at corresponding
column masses.
There is a rapid temperature increase in 
the transition region at $\log_{10}m_c\is -5.44$.
A temperature
plateau has been introduced at a temperature of 24~kK in order to get a Lyman-$\alpha$ flux similar
to the observed flux. 

\begin{figure}[hbtp]
  \includegraphics[width=\columnwidth]{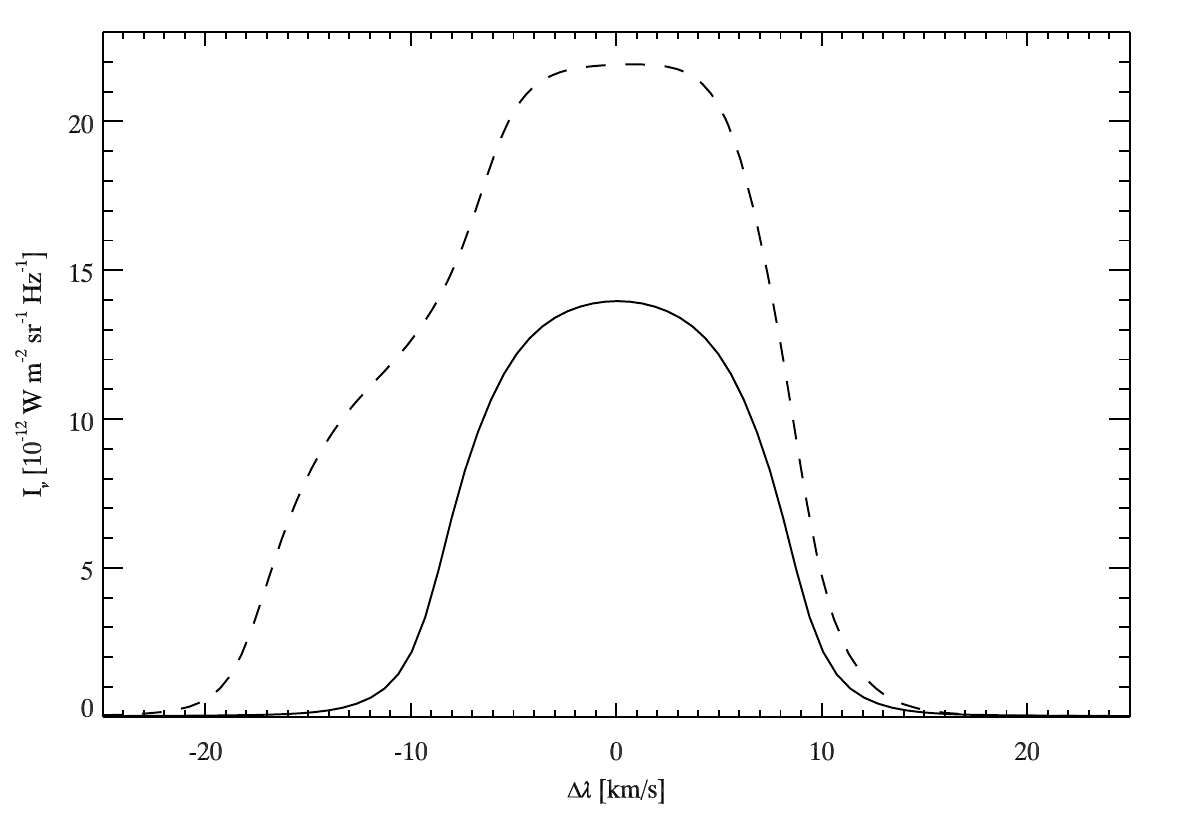} 
  \caption[]{\label{fig:i_lites} 
 Intensity as function of wavelength (in equivalent velocity units) for the \ciia\ line (solid) and the \ciib\ line (dashed) in the sunspot umbral model of \citet{1982ApJS...49..293L}.
 }
 \end{figure}

The emergent intensity for the \ciib\ lines is shown in Figure~\ref{fig:i_lites}. 
The ratio of the maximum intensities of the two lines is 1.6. 
The line profiles deviate from a Gaussian shape in
that the core part is flatter (especially for the stronger \ciib\ line) and the flanks are steeper than the best-fit Gaussian. The FWHM
of the intensity profile of the unblended \ciia\ line is 16.4 km s$^{-1}$. The temperature at the formation height is 24~kK and the 
microturbulence in the sunspot model is  2.6 km~s$^{-1}$. In an optically thin formation, the thermal width and the 
microturbulence would combine to give a FWHM of 10.5 km~s$^{-1}$. We thus have an opacity broadening factor of 1.6.

%===========================================================================
%===========================================================================
%%%%%%%%%%%%%%%%%%%%%%%%%%%%%%%%%%%%%

\begin{figure}[hbtp]
  \includegraphics[width=\columnwidth]{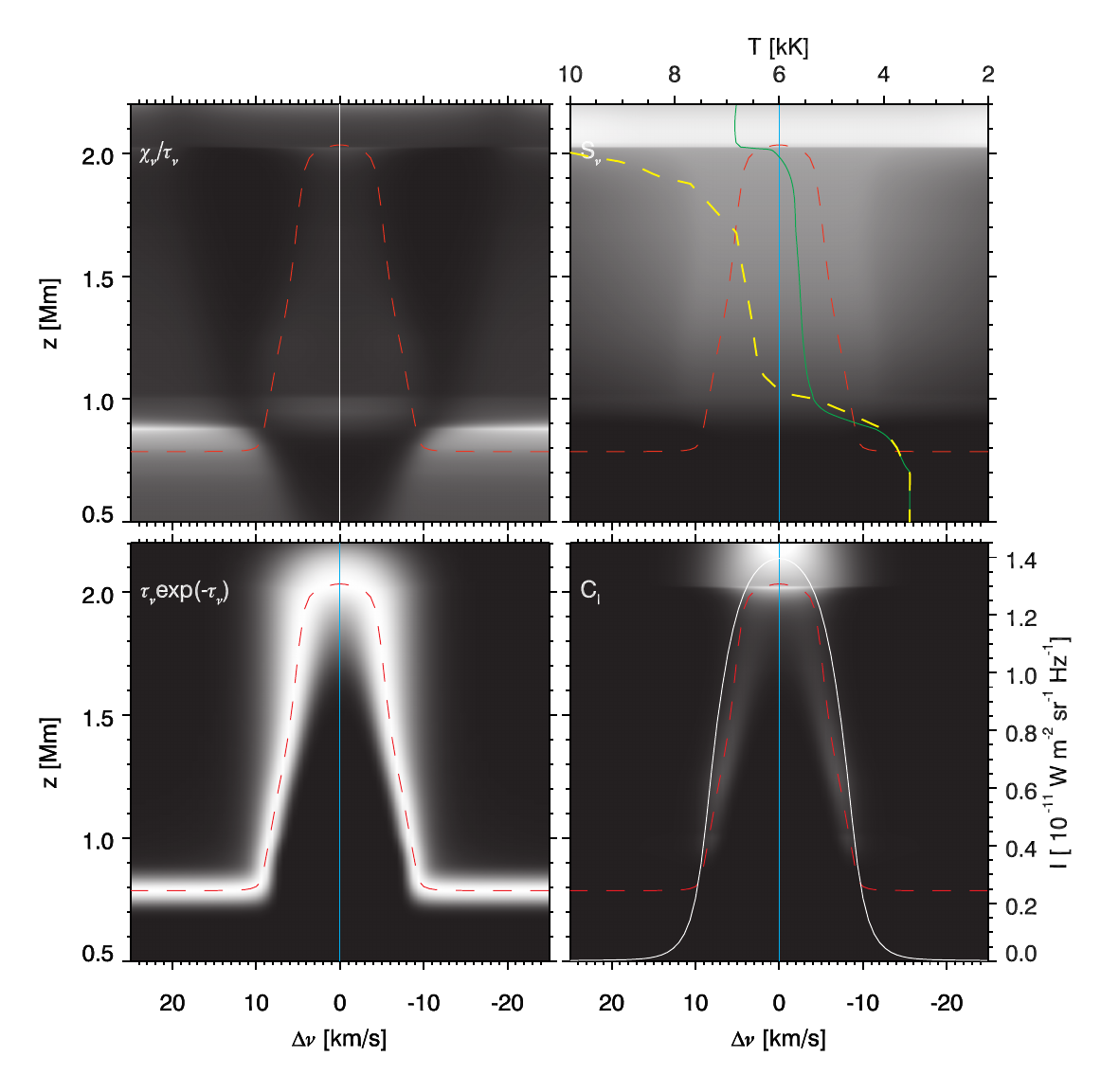} 
  \caption[]{\label{fig:sp1} 
   As Figure~\ref{fig:cia} but for the sunspot model atmosphere of \citet{1982ApJS...49..293L}. 
   }
\end{figure}

The intensity formation of the \ciia\ line in this atmosphere is shown in  Figure~\ref{fig:sp1}. 
The velocities for this atmosphere are set to zero.  The $\tau_{\nu}=1$ height (red dashed line in all panels) ranges from
0.8~Mm in the continuum to 2.0~Mm at the line center. 
The source function (solid green in upper right panel) is equal to the 
Planck function (yellow dashed line in the upper right panel) from the photosphere up to 0.9~Mm height
where it decouples. The source function continuously increase with the increase in temperature and becomes
maximum in the transition region. 
The monotonically increasing source function leads to a single peak emission line (white curve
in the lower right panel). The contribution function to intensity (image in lower right panel) is concentrated 
around 2.1 Mm height around optical depth unity which is also the lower part of the plateau at 24 kK in the model.
Much of the line intensity is thus formed at the temperature plateau which accounts for the rather flat line core. This 
is even more pronounced for the stronger \ciib\ line where a significant fraction of the line core is formed at the 
temperature plateau where the source function is rather constant (upper left panel).
%%%%%%%%%%%%%%%%%%%%%%%%%%%%%%%%%%%%%
%===========================================================================
%===========================================================================
%%%%%%%%%%%%%%%%%%%%%%%%%%%%%%%%%%%%%

\section{Discussion and Conclusions}\label{sec:cd}

% quintessential atomic model
The ionization balance and \edt{intensity} profiles of the \cii\ lines at 133.4--133.6 nm can be modelled with a 9-level 
model atom assuming complete frequency redistribution (CRD). 
% ionization balance
The ionization balance \ci/\cii\ is dominated
by photoionization and radiative recombination. The balance \cii/\ciii\ is dominated by collisional ionization
and dielectronic recombination. In the VAL3C atmosphere, carbon is more than 50\% in the 
form of \cii\ in the temperature range 5.6--35~kK, where especially the low temperature end is atmosphere 
dependent since it is set by photoionization which is a non-local process. 
% 3D scattering
Horizontal scattering is important
for the line core intensity which necessitates 3D non-LTE modelling and makes line core diagnostics 
influenced by non-local conditions.

% formation height
The line core is formed at temperatures 6--25~kK at a column mass close to 10$^{-6}$~g~cm$^{-2}$. The
continuum formation is at 300--800~km and is dominated by background opacity from neutral silicon. 
This background opacity needs to be treated in non-LTE in order to get the continuum intensity right but 
assuming LTE only affects the continuum intensity and not the line core intensity. The lower continuum
formation height is for kilo-Gauss flux concentrations in the photosphere.

% line width
The atomic absorption profile width is set by thermal and non-thermal broadening. The emergent intensity 
profile is broader than the atomic absorption profile because of the optical thick formation. This excess "opacity
broadening" is a factor of 1.2--4. The smaller opacity broadening happens for single peak intensity profiles 
that originate from a source function with a steep increase into the transition region and with a low 
chromospheric temperature. The larger opacity broadening is for double peak intensity profiles with central 
reversals that result from a source function with a local maximum  in the lower chromosphere because of a
steep temperature rise there.

% peak ratio
The intensity ratio between the \ciib\ and the \ciia\ line is lower than the optically thin 
value of 1.8 in the VAL3C model, the umbral model as well as for most columns in the \bifrost\ snapshot.
The optically thick formation can give any ratio, depending on the ratio of the source
functions of the two lines. 
The ratio is lower for double peak profiles where the intensity peaks are formed in the lower
chromosphere than for single peak profiles formed in the transition region.

% optically thin formation
The \cii\ lines are formed in the optically thick regime and erroneously assuming an optically thin formation
using the coronal approximation (neglecting photoionization) leads to too high 
formation temperatures and too low intensities.

In the next paper in this series we will discuss statistical correlations between atmospheric parameters
and observables and explore the diagnostic potential of the \cii\ lines.

\begin{acknowledgements}
The research leading to these results has received funding from the European Research Council under the
 European Union's Seventh Framework Programme (FP7/2007-2013) / ERC grant agreement no 291058.
This research was supported by the Research Council of Norway through the grant ``Solar Atmospheric 
Modelling'' and through grants of computing time from the Programme for Supercomputing. Bart De Pontieu 
and Jorrit Leenaarts are thanked for valuable comments on the manuscript.
\end{acknowledgements}

\bibliographystyle{apj}   
%\bibliography{CII_paper1}        

\end{document}